\documentclass{astlb}

\usepackage{graphicx}
\usepackage{color}

\usepackage{apjfonts}

\usepackage{multirow}
\usepackage[hyperfootnotes=false,draft]{hyperref}

\definecolor{darkblue}{rgb}{0,0,0.9}

\def\doubleline{\vskip 3pt\hrule \vskip 1.5pt \hrule \vskip 5pt}
\def\neff{\ensuremath{N_{\mathrm{eff}}}}

\def\s8data{$D(\sigma_8)$}

\begin{document}

\journalinfo{2018}{0}{0}{1}[0]
%\UDK{524.77}

\title{The measurements of matter density perturbations amplitude\\
  from cosmological data}

\author{R.A.~Burenin\email{rodion@hea.iki.rssi.ru}\\
  \vskip -2mm ~\\
  \emph{Space Research Institute of RAS (IKI), Moscow, Russia}
}

\shortauthor{R. A. Burenin}

\shorttitle{The measurements of matter density perturbations amplitude}

%\submitted{\today}
\submitted{May 4, 2018}

\begin{abstract}
  We compare various physically different measurements of linear
  matter density perturbation amplitude, $sigma_8$, which are obtained
  from the observations of CMB anisotropy, galaxy cluster mass
  function, weak gravitation lensing, matter power spectrum and
  redshift space distortions. We show that $\sigma_8$ measurement from
  CMB gravitational lensing signal based on \emph{Planck} CMB
  temperature anisotropy data at high multipoles, $\ell>1000$,
  contradict to all other measurements obtained both from remaining
  \emph{Planck} CMB anisotropy data and from other cosmological data,
  at about $3.7\sigma$ significance level. Therefore, these data
  currently \emph{should not} be combined with other data to constrain
  cosmological parameters.
  
  With the exception of \emph{Planck} CMB temperature anisotropy data
  at high multipoles, all other measurements are in good agreement
  between each other and give the following measurements of linear
  density perturbation amplitude: $\sigma_8=0.792\pm0.006$, mean
  density of the Universe: $\Omega_m=0.287\pm0.007$, and Hubble
  constant: $H_0 = 69.4\pm 0.6$~km~s$^{-1}$~Mpc$^{-1}$. Taking in
  account the data on baryon acoustic oscillations and (or) direct
  measurements of the Hubble constant, one can obtain different
  constraints on sum of neutrino mass and number of relativistic
  species.
  
  
\end{abstract}

\section{Introduction}

The measurements of cosmic microwave background (CMB) temperature and
polarization obtained in \emph{Planck} all-sky survey
\citep{planck15_rew,planck15_cosm}, constitute one of the basic
cosmological datasets, currently used to constrain parameters of
cosmological model. However, it is known that there is a number of
tensions between cosmological parameters constraints, which are
obtained from different subsets of \emph{Planck} survey data as well
from the data of some other cosmological measurements. These tensions
were discussed previously in many papers
\citep[e.g.,][]{addison16,planck15_like,planck15_cosm,couchot17,planck15_par_shifts}.
For example, the constraints on the linear matter density
perturbations amplitude perturbations obtained from measurements of
galaxy cluster mass function \citep[e.g.,][]{av09a,av09b,PSZ2cosm},
and also from weak lensing measurements
\citep{heymans13,troxel17,des_gwl_y1_17}, appears to be lower than the
constraints obtained from CMB temperature anisotropy in \emph{Planck}
survey.

\begin{figure*}
  \centering
  \includegraphics[width=0.33\linewidth]{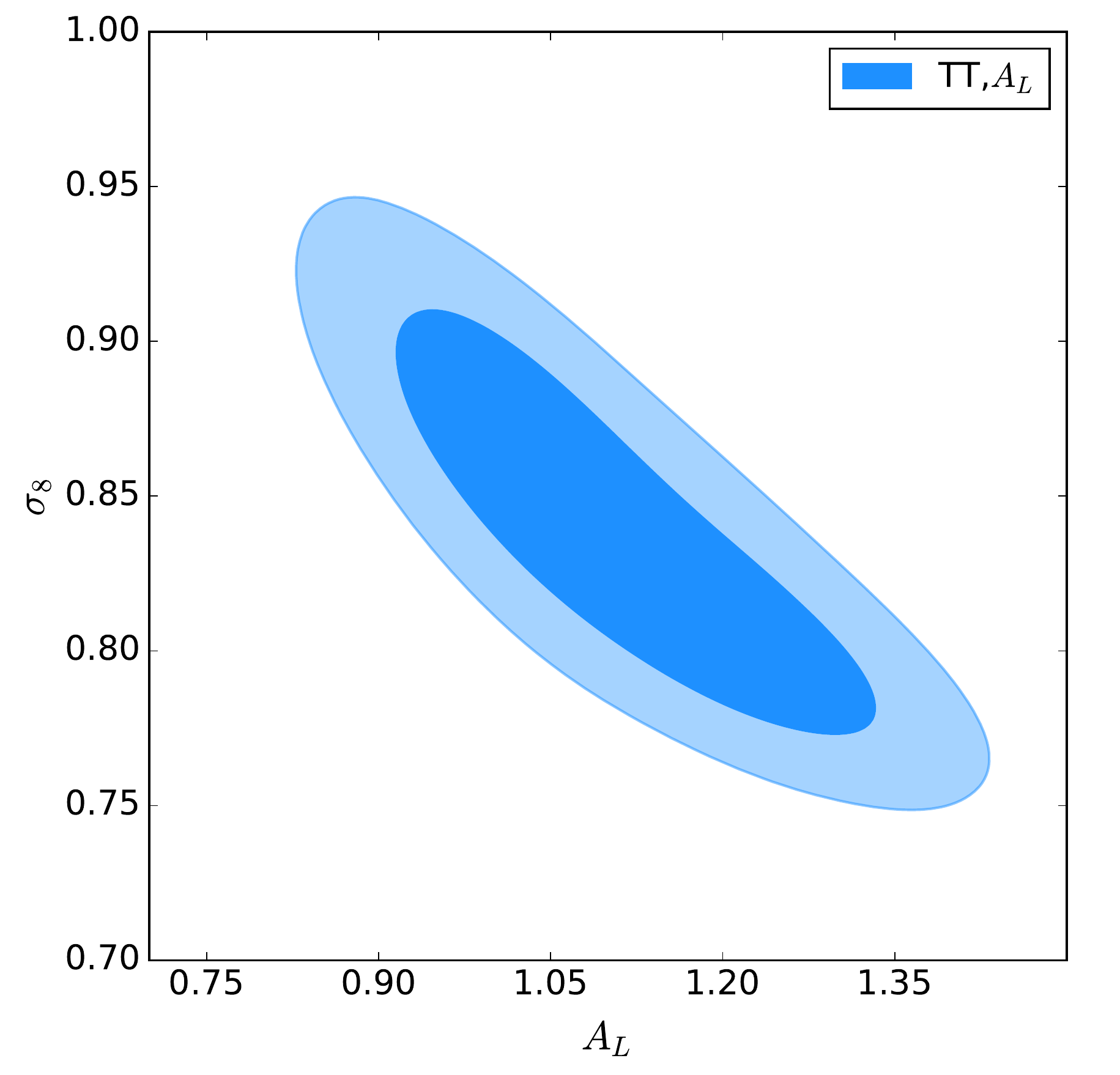}
  ~~
  \includegraphics[width=0.33\linewidth]{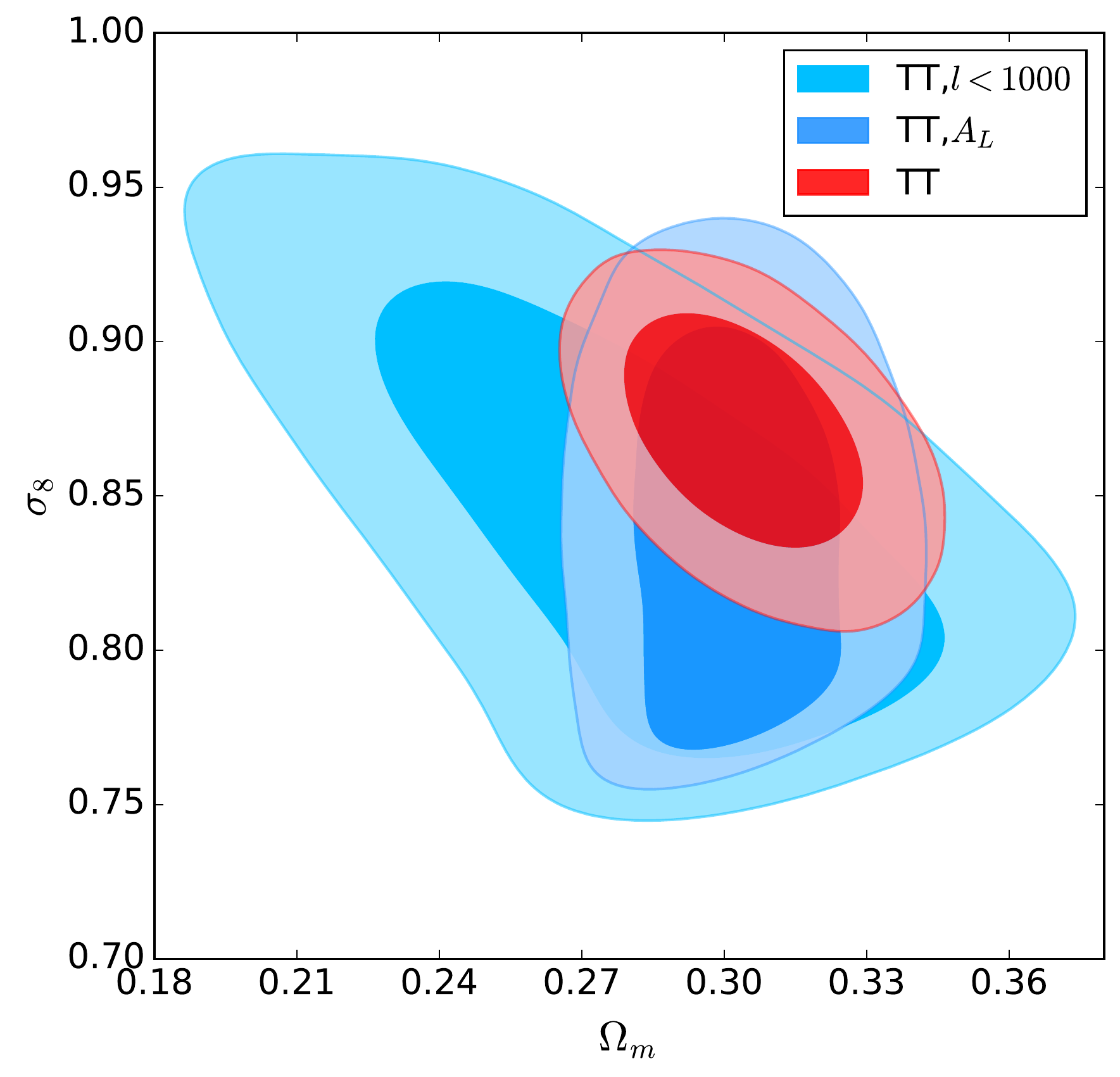} 
  
  \caption{The constraints on gravitational lensing amplitude, $A_L$,
    mean matter density, $\Omega_m$, and linear density perturbations
    amplitude, $\sigma_8$, from \emph{Planck} survey CMB temperature
    anisotropy spectrum data. Different contours show the constraints
    from the CMB temperature anisotropy spectrum in
    \emph{$\Lambda$CDM} model (\emph{TT}), in the same model with free
    lensing amplitude (\emph{TT},$A_L$), and also in
    \emph{$\Lambda$CDM} model from \emph{Planck} CMB temperature
    anisotropy spectrum at multipoles below $\ell=1000$
    (\emph{TT},$\ell<1000$).}
  \label{fig:alens}
\end{figure*}

Below we compare various constraints on linear matter density
perturbations amplitude, obtained using various observables and
various cosmological data such as measurements of CMB anisotropy,
galaxy cluster mass function, weak gravitation lensing, matter power
spectrum and redshift space distortions. It is shown that the
cosmological parameters constraints obtained using these heterogeneous
data are in good agreement with each other, with the exception of the
data on \emph{Planck} CMB temperature anisotropy spectrum at high
multipoles, $\ell>1000$. The cosmological parameters constraints,
which can be obtained with no use of this part of \emph{Planck} survey
data, are studied in detail.

The calculations of the cosmological parameters constraints were
carried out by Markov Chains Monte-Carlo (MCMC) simulations using
\emph{CAMB} \citep{lewis00} and \emph{CosmoMC} software
\citep{lewis02,lewis13}, version of November 2016. The \emph{CosmoMC}
software was modified to include the galaxy cluster mass function data
from \cite{av09a,av09b}, the latest South Pole Telescope CMB
polarization data \citep{henning18}, Dark Energy Survey weak lensing
data \citep{des_gwl_y1_17}, and the most recent baryon acoustic
oscillations (BAO) data from Sloan Digital Sky Survey
\citep{alam17}. In all MCMC simulations cosmological model was defined
in the same way, as it was done in \cite{planck15_cosm}. In all
figures below the contours at 68\% and 95\% confidence level are
shown.  All the numerical values of the confidence intervals are given
at 68\% confidence level, upper limits --- at 95\% confidence level.

\section{\emph{Planck} CMB anisotropy data}

CMB anisotropy data allow to measure the density perturbation
amplitude in several ways. The constraints on, $\sigma_8$ --- linear
density perturbation amplitude at 8~Mpc scale, obtained from
\emph{Planck} CMB temperature anisotropy data
\citep[\emph{TT},][]{planck15_like,planck15_cosm}, are shown in
Fig.~\ref{fig:alens} with red contours. This constraint is obtained
mainly from the measurement of the amplitude of CMB gravitation
lensing on large scale structure. It can be easily shown, for example,
as follows. If one thaw the phenomenological lensing amplitude $A_L$
(the coefficient by which the lensing potential, $C^{\phi\phi}$, is
multiplied before the calculation of temperature anisotropy lensing is
done, see, e.g., \citealt{planck15_cosm}), it appears to be well
correlated with density perturbation amplitude $\sigma_8$ (see
Fig.~\ref{fig:alens}, left panel). For higher values of $A_L$ the data
require to lower perturbations amplitude $\sigma_8$ so that measured
lensing amplitude of CMB temperature anisotropy is the same.  The best
$\sigma_8$ constraint is obtained under the condition $A_L=1$, when
the measured lensing amplitude is directly converted into density
perturbations amplitude.

The gravitational lensing of CMB temperature anisotropy on large scale
structure is observed as smoothing of acoustic peaks at high
multipoles, approximately at $\ell>1000$ \citep[see,
e.g.,][]{planck15_cosm}. If CMB temperature anisotropy spectrum is
truncated at those high multipoles, only much weaker $\sigma_8$
constraint can be obtained, as expected (see Fig.~\ref{fig:alens},
right panel).

\begin{figure*}
  \centering
  \includegraphics[width=0.33\linewidth]{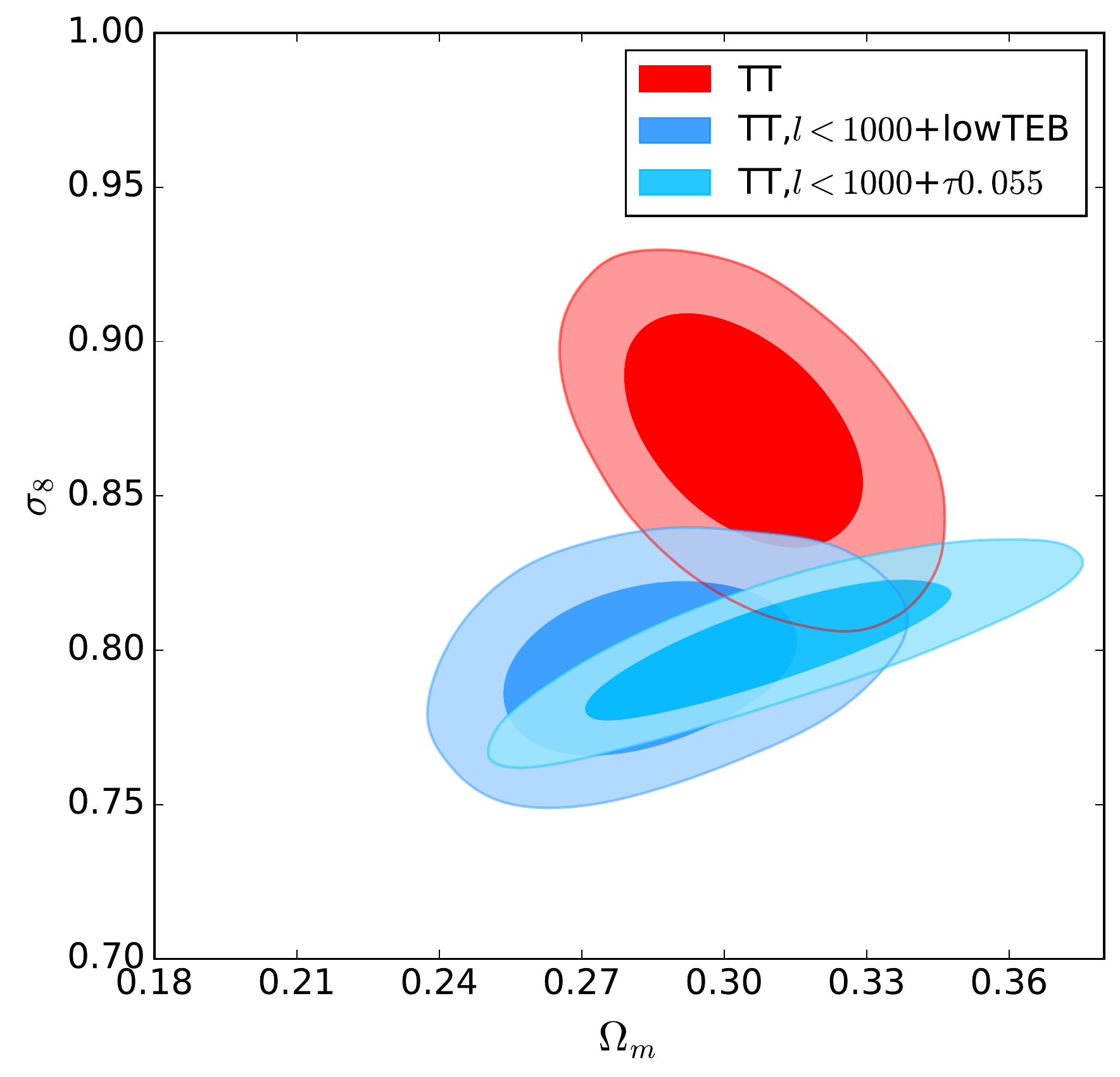} 
  \includegraphics[width=0.33\linewidth]{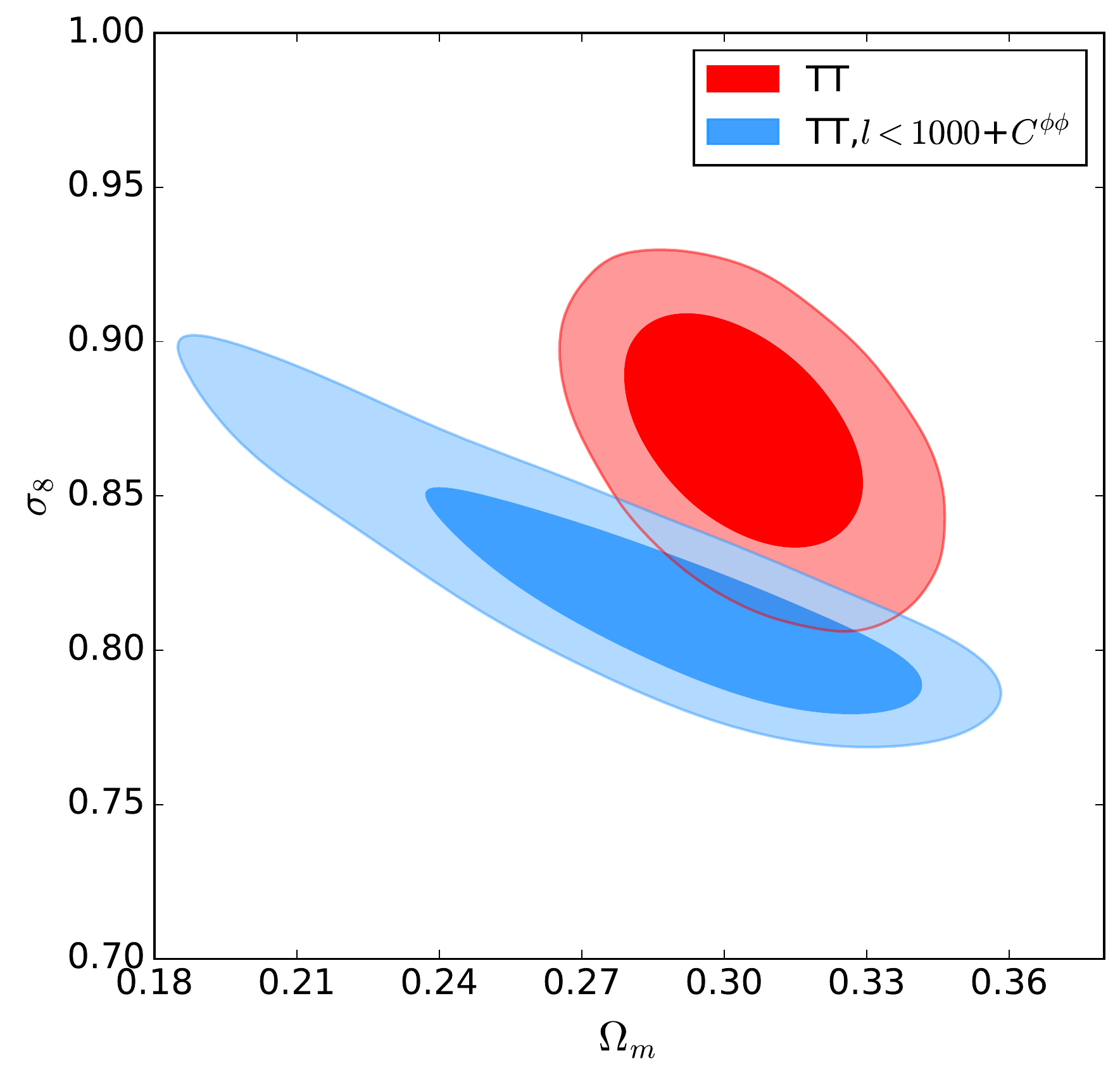}
  \includegraphics[width=0.33\linewidth]{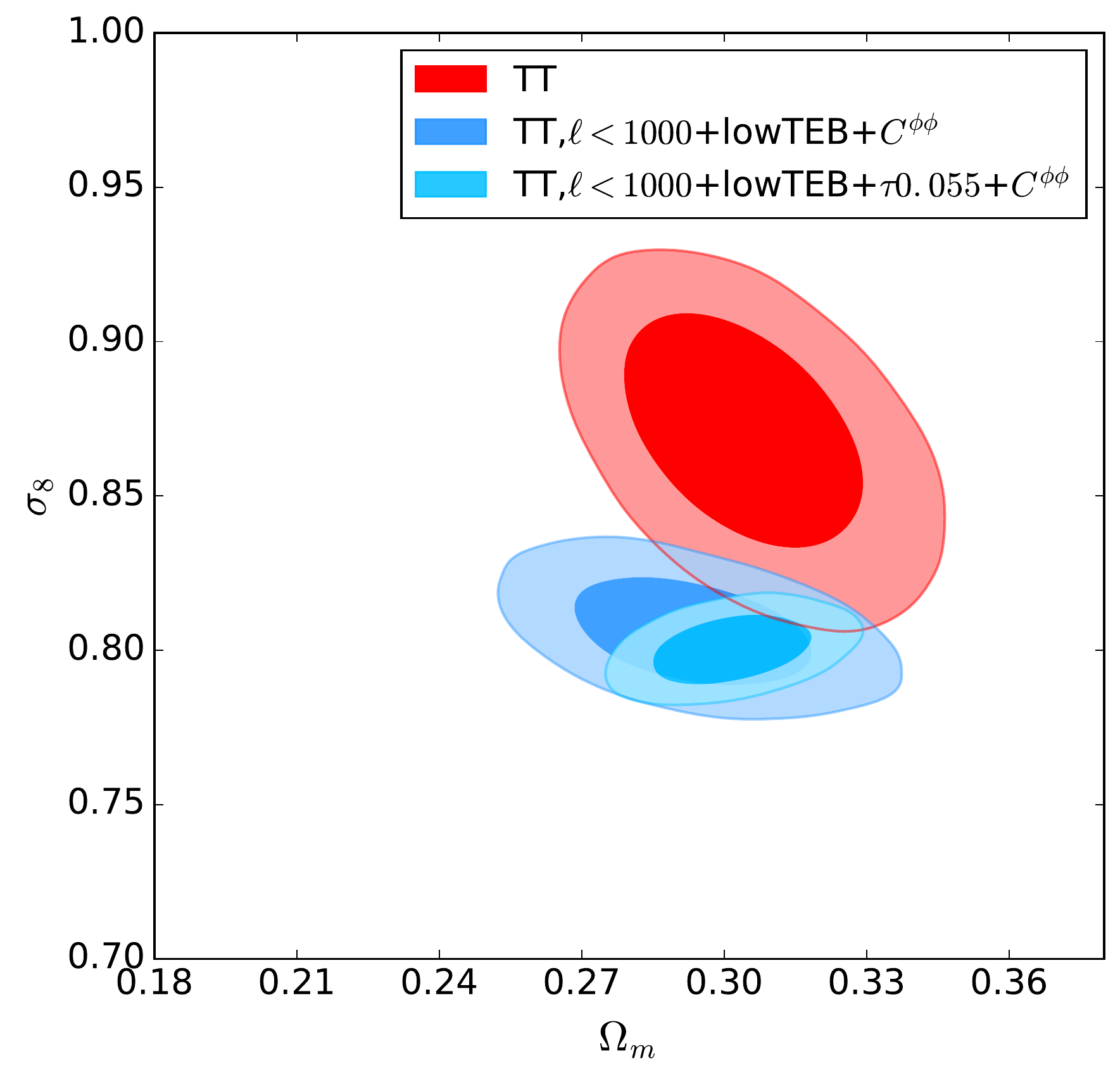}

  \caption{The constraints on mean matter density, $\Omega_m$, and
    linear density perturbations amplitude, $\sigma_8$, in
    \emph{$\Lambda$CDM} model, from the \emph{Planck} CMB temperature
    anisotropy spectrum (\emph{TT}, red contours), and also from the
    data on CMB polarization at low multipoles (\emph{lowTEB},
    $\tau0.055$) and lensing potential ($C^{\phi\phi}$), in
    combination with \emph{Planck} CMB temperature anisotropy data at
    $\ell<1000$.}
  \label{fig:s8-om-pl} 
\end{figure*}

The exact position of the lower boundary of high multipole range in
\emph{Planck} CMB anisotropy spectrum, $\ell=1000$, is more or less
arbitrary. It is justified by the fact that this value allow to select
the region where the effect of gravitational lensing is most
prominent, this value approximately corresponds to the minimum between
third and fourth acoustic peaks, and also this number is a round one,
i.e.\ this choice emphasize that no any adjustment of this value was
made. Note, that similar definition of high multipole region in
\emph{Planck} CMB anisotropy spectrum was used in some other works
\citep{addison16,henning18}, although with somewhat different
justification.

Note that the constraints on lensing amplitude $A_L$ obtained from
\emph{TT} data only, allows both standard value $A_L=1$ and also
higher values, $A_L\approx1.2$ (see Fig.~\ref{fig:alens}, left
panel). These data are typically used in combination with other data,
e.g., with CMB polarization measurements at low multipoles, which
produce lower value of perturbations amplitude $\sigma_8$ (see
below). This explains, why the higher values of $A_L\approx1.2$ were
measured in some earlier works, which was fairly considered as anomaly
\citep[e.g.,][]{planck15_cosm,planck15_par_shifts,couchot17}.

From the data on the CMB polarization at low multipoles one can obtain
the constraint on reionization optical depth, $\tau$, which allow to
constrain density perturbation amplitude using the measurement of the
CMB temperature anisotropy amplitude itself. In \emph{Planck} 2015
data release the CMB polarization data obtained with Low Frequency
Instrument (LFI) were published \citep{planck15_lfi,planck15_cosm},
these data are included in \emph{Planck} likelihood software
(\emph{lowTEB}). In addition, somewhat later CMB polarization data at
low multipoles obtained with High Frequency Instrument (HFI) were
published \citep{planck_hfi_tau1,planck_hfi_tau2}. The likelihood for
these data (\emph{SimLow}) is used below as a prior for reionization
optical depth $\tau=0.055\pm0.009$ \citep{planck_hfi_tau1}.

In order to do not mix the constrains from this and other individual
dataset with the constraints from CMB temperature anisotropy spectrum
at high multipoles discussed above, the calculations all of the
constraints from individual datasets below are done in combination
with \emph{Planck} TT data at multipoles $\ell<1000$. Resulting
constraints from \emph{Planck} CMB polarization anisotropy data are
presented in the left panel of Fig.~\ref{fig:s8-om-pl}. One can see
the notable tension of these constraints with those from the complete
\emph{Planck} TT spectrum, where the constraint on $\sigma_8$ is
produced mainly by the data at high multipoles, as it was discussed
above.

The density perturbations amplitude can also be constrained directly
from the amplitude of observed lensing potential spectrum,
$C^{\phi\phi}$, which can be recovered from four-point statistics
(trispectrum) which was also done using \emph{Planck} survey data
\citep{planck15_lens}. The constrains on $\Omega_m$ and $\sigma_8$,
obtained with these data are shown in the middle panel of
Fig.~\ref{fig:s8-om-pl}. These data are also in some tension with
\emph{Planck} TT data at high multipoles.

The $\Omega_m$ and $\sigma_8$ constraints obtained from the combined
CMB polarization data at low multipoles and CMB lensing potential data
are shown in the right panel of Fig.~\ref{fig:s8-om-pl}. These data,
taken together, give the constraint $\sigma_8 = 0.800\pm 0.007$, while
\emph{Planck} TT data give $\sigma_8 = 0.870\pm 0.025$. Therefore,
there is approximately $2.7\sigma$ tension between these datasets.

%Note, that the tension

In fact, it is this tension in measurements of density perturbation
amplitude is the source of well known anomaly in \emph{Planck} survey
data related to the observation of increased lensing amplitude $A_L$,
which was extensively discussed earlier
\citep[e.g.,][]{planck15_cosm,planck15_par_shifts,couchot17}. As it
have been discussed above, higher value of $A_L$ allows to measure
lower value of $\sigma_8$, which allow to reconcile the constraints
presented in Fig.~\ref{fig:s8-om-pl}. From this figure one can also
see that new \emph{Planck} HFI data on CMB polarization at low
multipoles should enhance this anomaly, which was also observed
\citep{planck_hfi_tau1}.

Some of these internal tensions in Planck CMB data were noted and
discussed in earlier works. The tensions between the parameters of
\emph{$\Lambda$CDM} model obtained from CMB temperature anisotropy
data at multipoles below and above $\ell=1000$ were discussed by
\cite{addison16}. These tensions were also discussed in Planck
Collaboration papers (XI, XIII, 2016; LI, 2017) where their
significance was found to be not compelling enough. Recently
\cite{motloch_hu_18} found that these tensions can be detected in
Planck CMB temperature spectrum at about 2.4$\sigma$ significance
independently from the cosmological model and they are driven by
smoothing of the acoustic peaks near multipoles
$\ell\approx 1250$--$1500$. This is in good agreement with the results
discussed above, since it is this extra smoothing of acoustic peaks
should be observed in result of the presence of extra gravitational
lensing signal in Planck CMB temperature anisotropy spectrum.

\begin{figure*}
  \centering
  \includegraphics[width=0.32\linewidth]{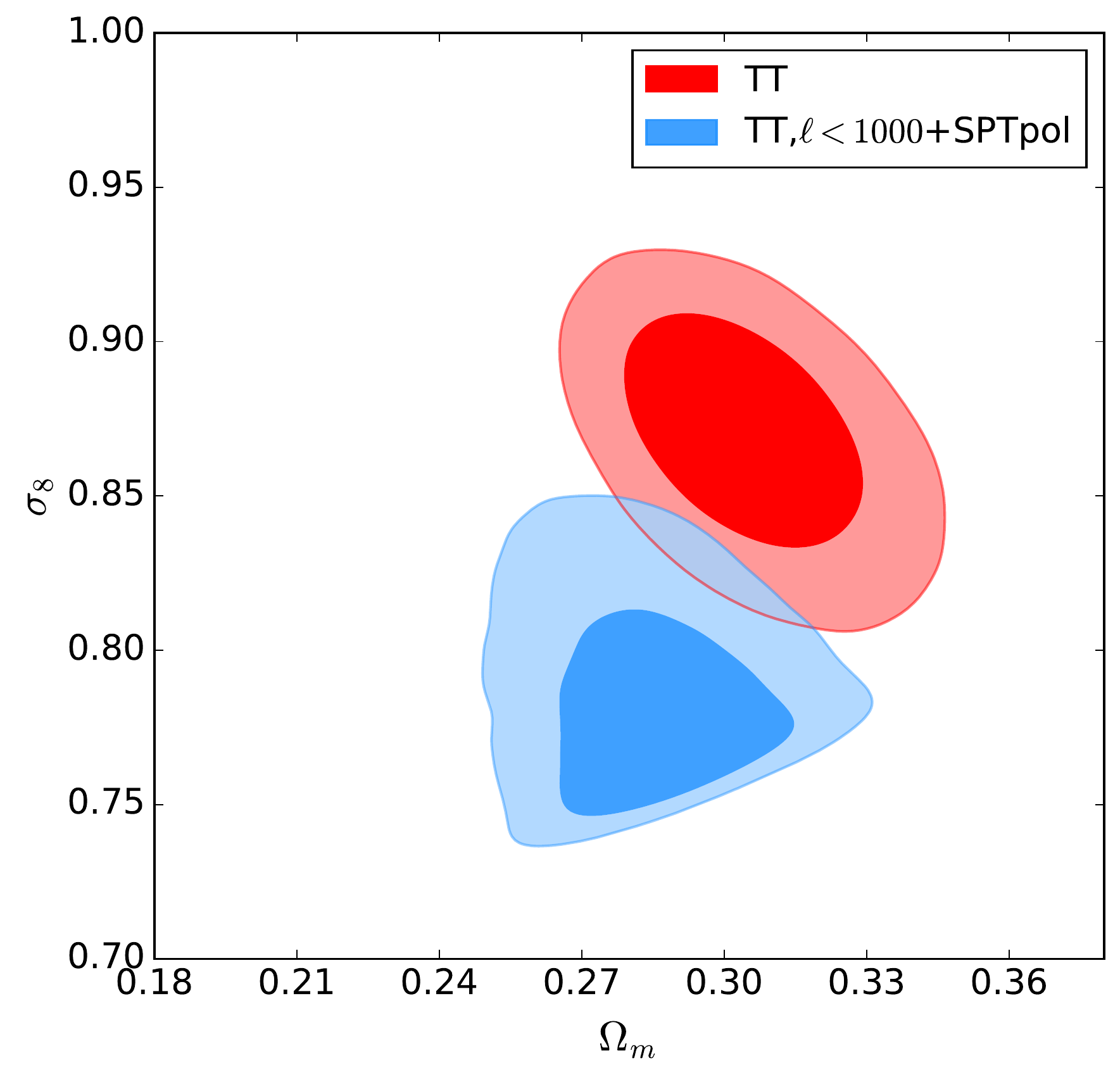}
  \includegraphics[width=0.32\linewidth]{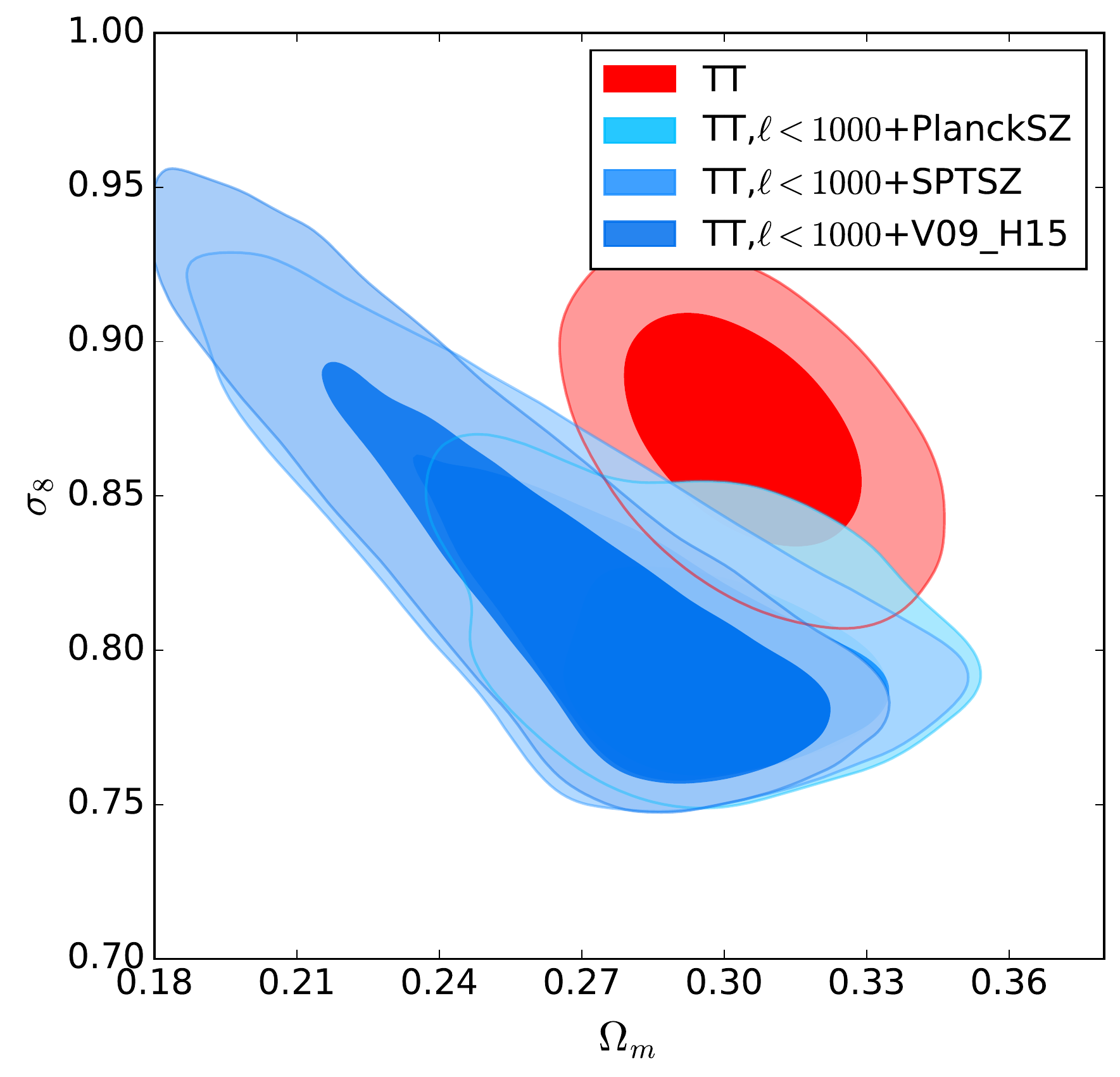} 
  
  \includegraphics[width=0.32\linewidth]{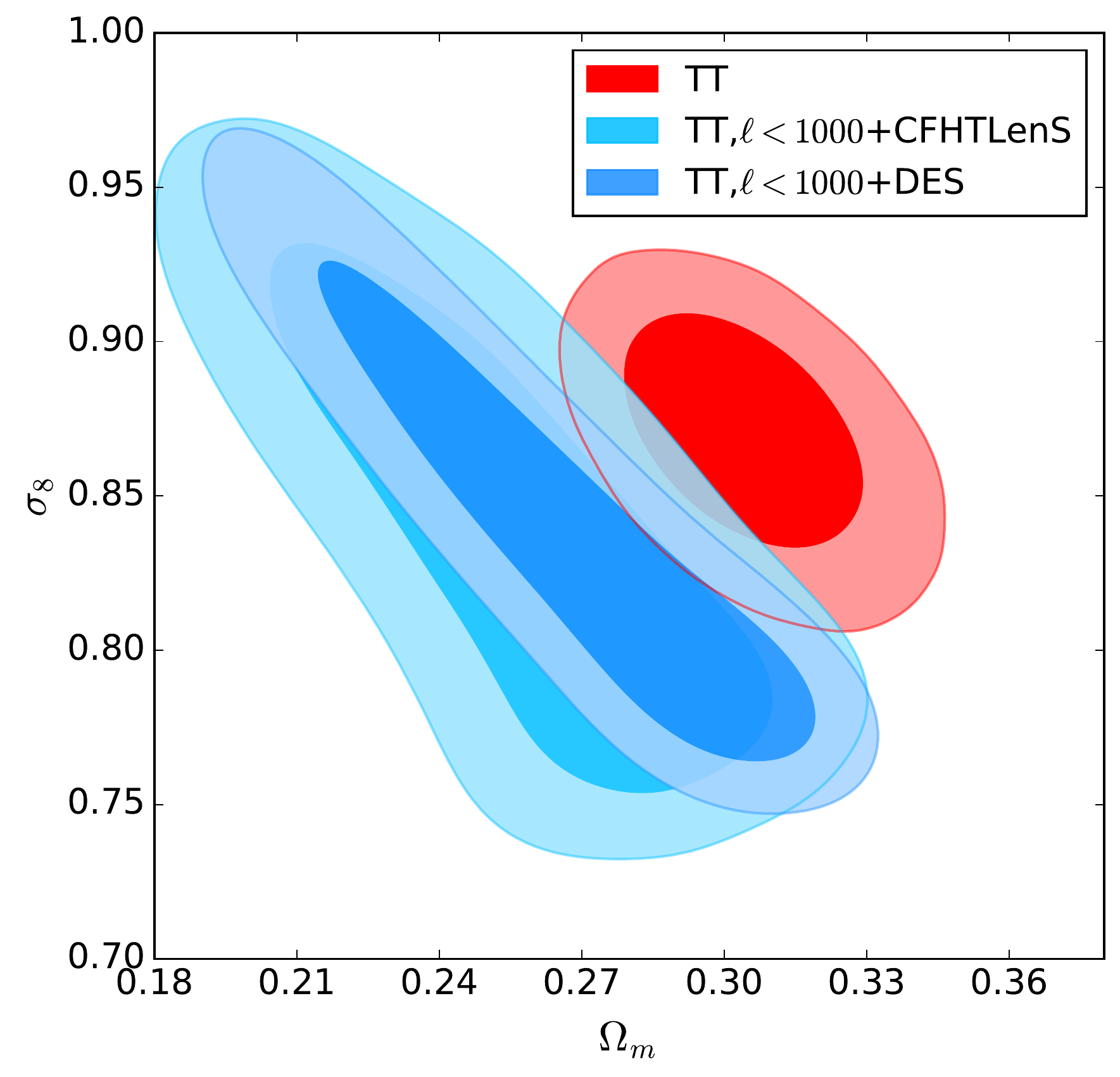}
  \includegraphics[width=0.32\linewidth]{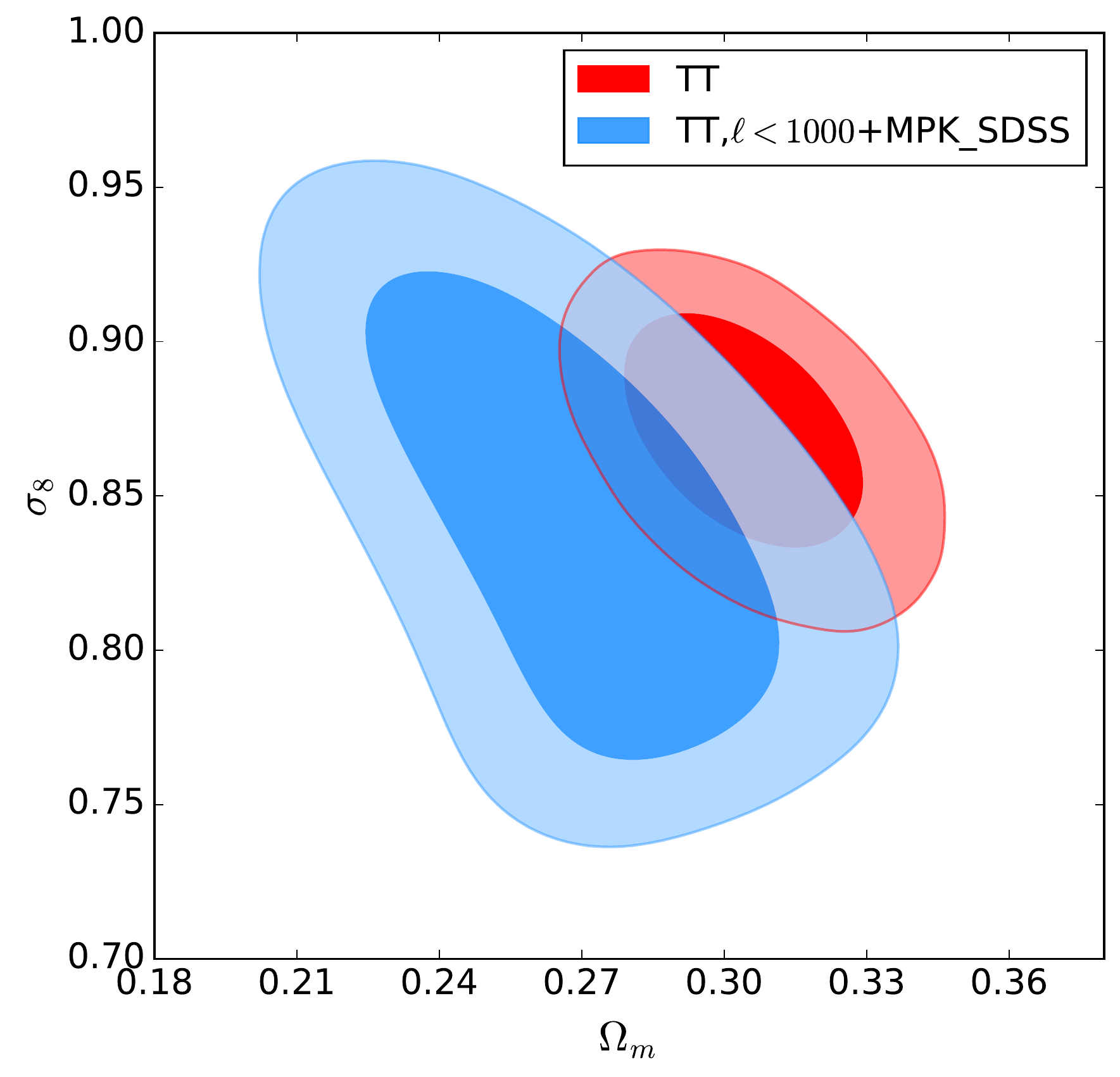}

  \includegraphics[width=0.32\linewidth]{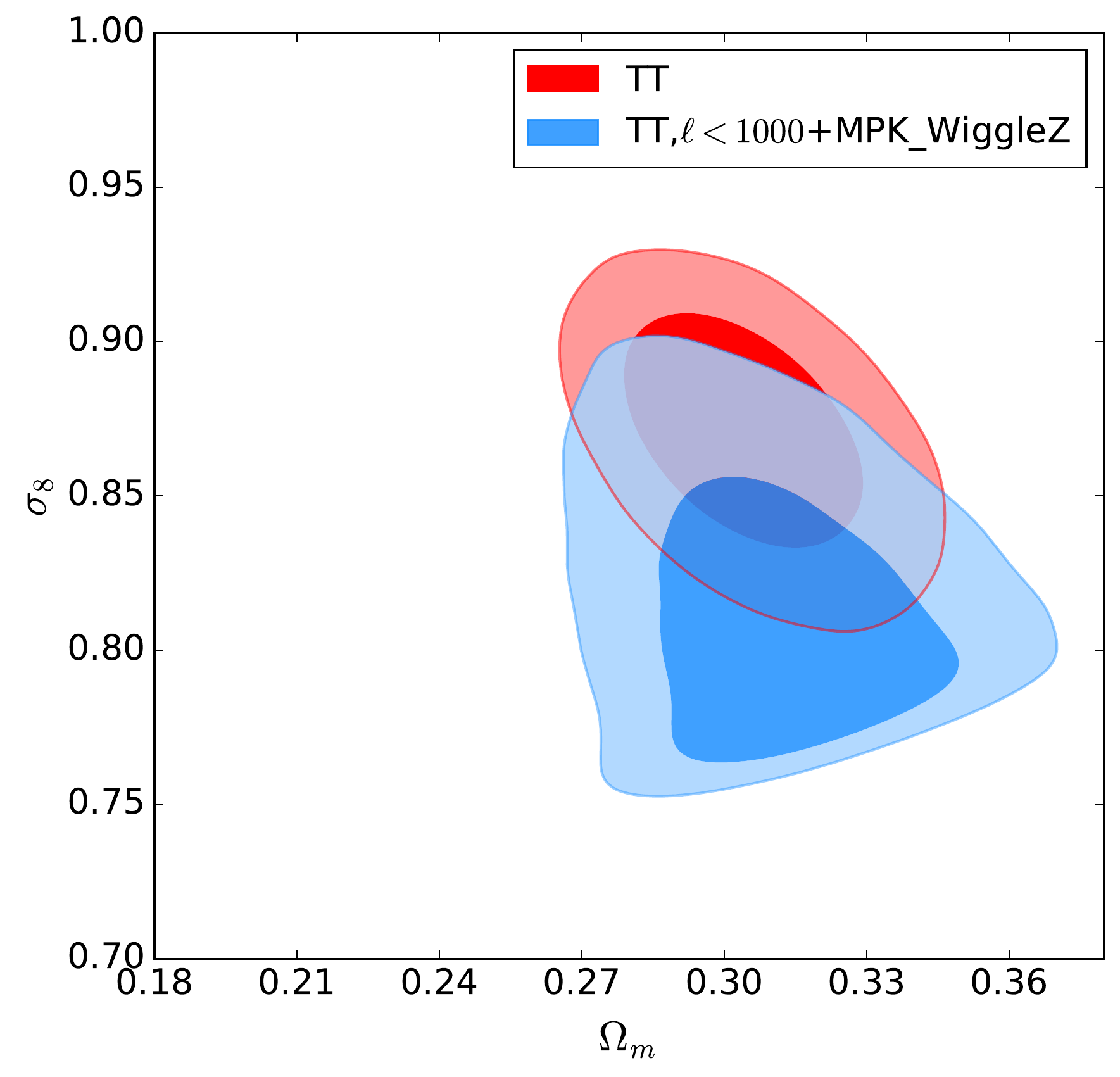}
  \includegraphics[width=0.32\linewidth]{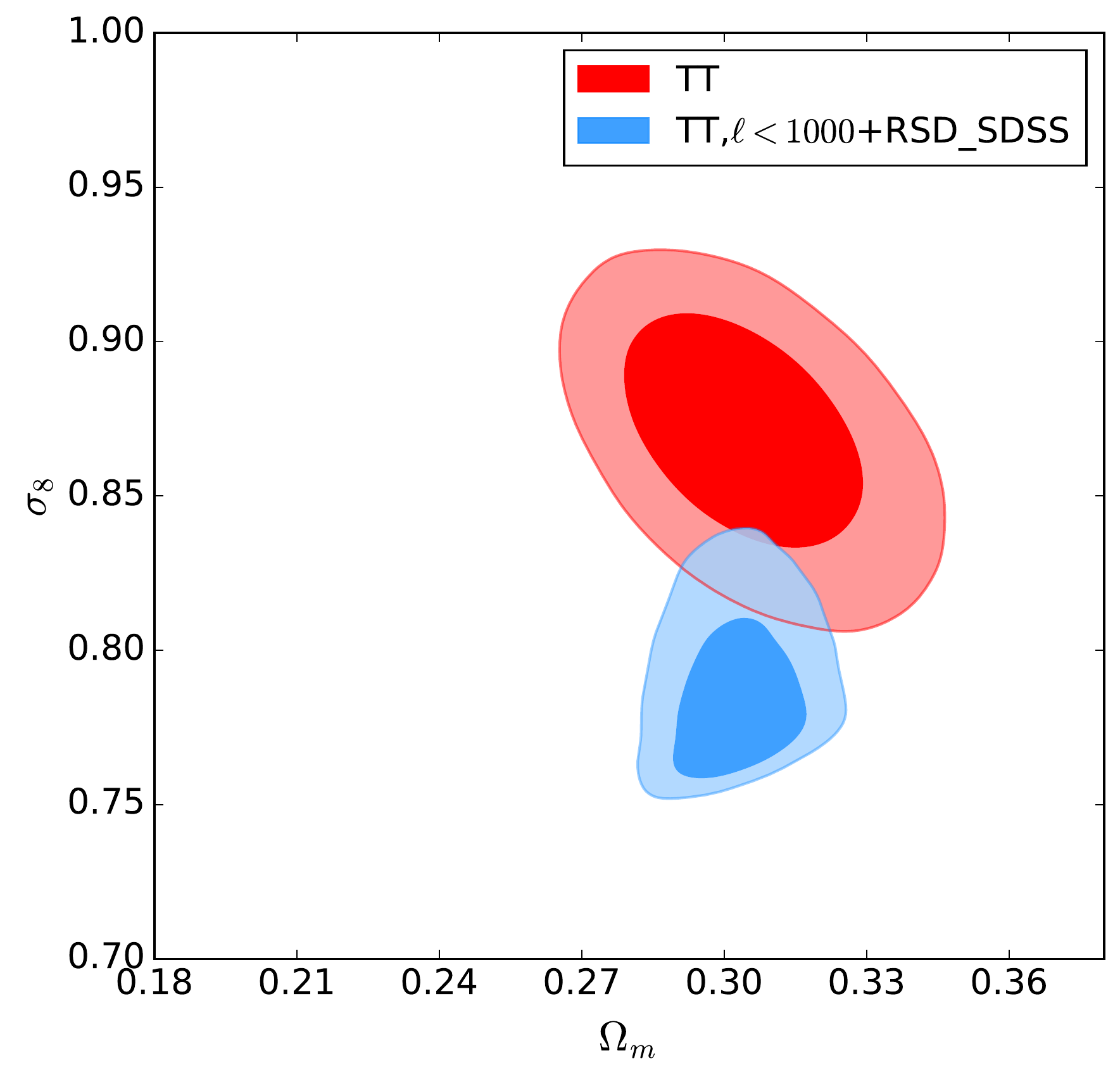}
  
  \caption{The constraints on mean matter density, $\Omega_m$, and
    linear density perturbations amplitude, $\sigma_8$, in
    \emph{$\Lambda$CDM} model from \emph{Planck} CMB temperature
    anisotropy spectrum (TT, red contours), and also from the
    following data: upper row, left panel --- South Pole Telescope CMB
    polarization anisotropy measurements (\emph{SPTpol}), right panel
    --- galaxy cluster mass function measurements (\emph{PlanckSZ,
      SPTSZ, V09\_H15}), middle row, left panel --- weak lensing of
    distant galaxies (\emph{CFHTLenS}, \emph{DES}), right panel ---
    power spectrum of galaxies in SDSS 4-th data release
    (\emph{MPK\_SDSS}), lower row, left panel --- power spectrum of
    galaxies in \emph{WiggleZ} survey (\emph{MPK\_WiggleZ}), right
    panel --- redshift space distortions measured in SDSS 12-th data
    release (\emph{RSD\_SDSS}). All these data are combined with
    \emph{Planck} CMB temperature anisotropy data at $\ell<1000$.}
  \label{fig:s8-om-all-sep} 
\end{figure*}

\begin{figure*}
  \centering
  \includegraphics[width=0.4\linewidth]{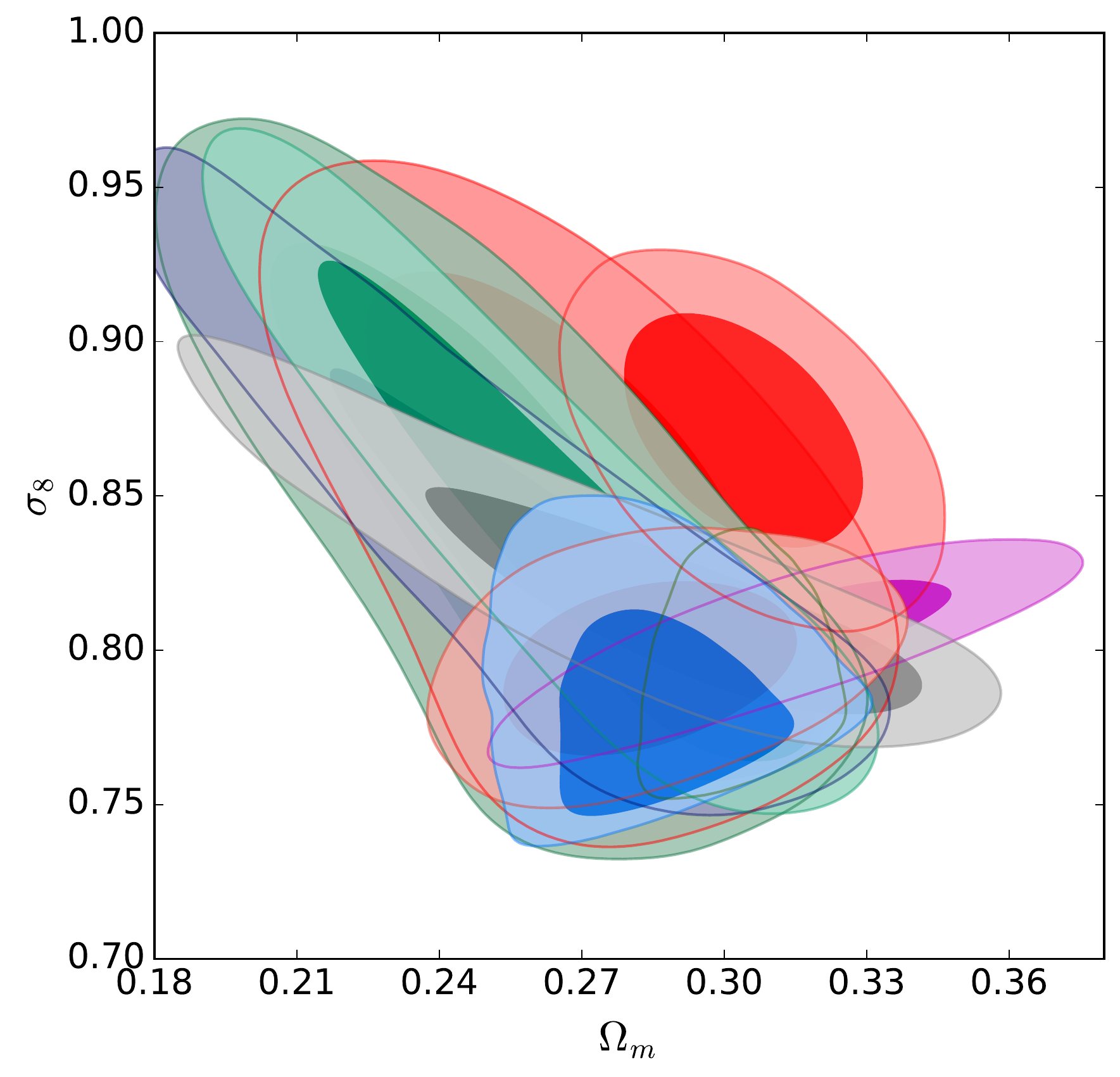} 
  \includegraphics[width=0.4\linewidth]{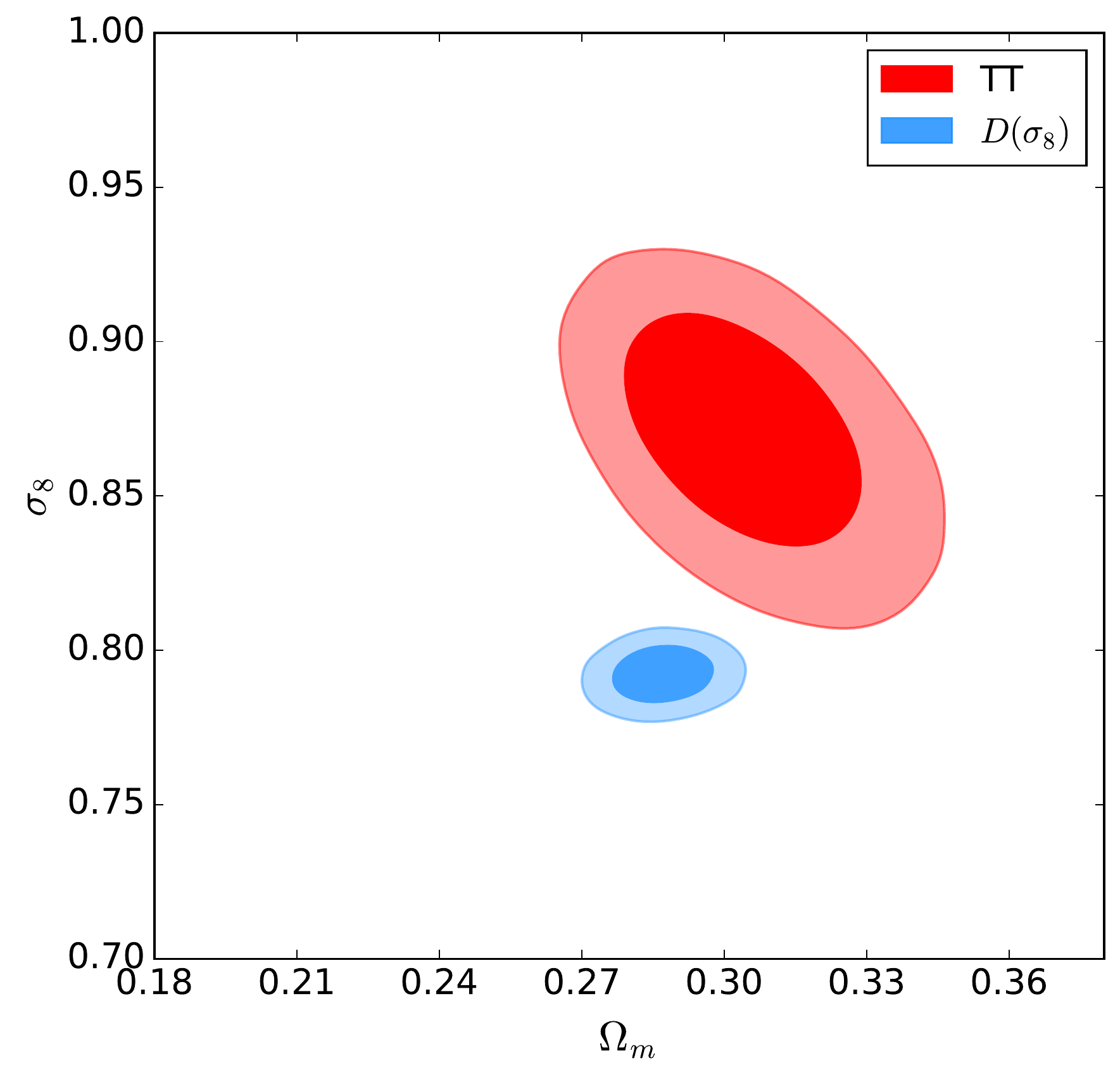}
  
  \caption{The constraints on mean matter density, $\Omega_m$, and
    linear density perturbations amplitude, $\sigma_8$, in
    \emph{$\Lambda$CDM} model from \emph{Planck} CMB temperature
    anisotropy spectrum data (TT, red contours), and also from all
    other data presented in Fig.~\ref{fig:s8-om-pl}, taken together
    (see text for details).}
  \label{fig:s8-om-all} 
\end{figure*}

\section{Other measurements of density perturbation amplitude }

\subsection{South Pole Telescope CMB polarization anisotropy data}

Gravitation lensing produce the smoothing of acoustic peaks not only
in the spectrum of CMB temperature anisotropy, but also in CMB
polarization spectrum. High sensitive measurements of CMB polarization
anisotropy were obtained recently using the data of new \emph{SPTpol}
detector at South Pole Telescope \citep{henning18}. The constraints on
mean matter density $\Omega_m$, and linear density perturbations
amplitude, $\sigma_8$, in \emph{$\Lambda$CDM} model, obtained with
these data are presented in Fig.~\ref{fig:s8-om-all-sep}, upper row,
left panel.

These data constrain $\sigma_8$ at notably lower values, as compared
to \emph{Planck} temperature anisotropy spectrum data (\emph{TT}, red
contours). This tension was discussed by \emph{SPTpol} collaboration
\citep{henning18}, who also note that it originates due to the
\emph{SPTPol} data at high multipoles, $\ell>1000$, while the data at
lower multipoles, $\ell<1000$, are found to be in agreement with
\emph{Planck} survey data. We note, that \emph{SPTPol} data both at
low and high multipoles taken separately, produce only weak constraint
on density perturbations amplitude. There is no good $\sigma_8$
constraint from the data at $\ell>1000$ since spectral index, $n_s$,
is poorly constrained with these data and is overestimated to the
values of up to $n_s\approx1.1$. More accurate $\sigma_8$ measurement
can be obtained only with \emph{SPTPol} data at all multipoles, which
are shown in Fig.~\ref{fig:s8-om-all-sep} and are used below.

\subsection{Galaxy clusters}

The measurements of galaxy clusters mass function give one of the most
sensitive methods to constrain matter density perturbations amplitude.
These measurements were obtained in many works, based on cluster
samples selected in X-rays \citep[e.g.,][]{av09a,av09b,mantz15},
optical \citep[e.g.,][]{rozo10}, and also using the observations of
Sunyaev-Zeldovich effect \citep[SZ,][]{sz72} in microwave band
\citep[e.g.,][]{dehaan16,PSZ2cosm}.

All these measurements produce approximately similar results, since
the accuracy of these measurements is limited by the uncertainty of
the measurements of cluster masses and they are based on similar
galaxy cluster mass scale calibrations. The main method to calibrate
galaxy cluster mass scale is based on the observations of weak
gravitational lensing in clusters, the results of these calibrations
are more or less consistent among each other
\citep[e.g.,][]{israel14,hoekstra15,smith16,applegate16,dietrich17}.

Fig.~\ref{fig:s8-om-all-sep}, upper row, right panel, shows the
constraints on linear density perturbations amplitude, $\sigma_8$,
obtained from the various galaxy cluster mass function datasets ---
the galaxy cluster data of \emph{Planck} SZ survey
\citep[\emph{PlanckSZ},][]{PSZ2cosm}, South Pole Telescope
\citep[\emph{SPTSZ},][]{dehaan16}, and also the data on clusters
selected in X-rays from \cite{av09a,av09b}, see also \cite{bv12}. In
the last case the constraints were corrected on the basis of recent
cluster mass scale calibrations \citep{hoekstra15}, and were taken in
the form: $\sigma_8 = 0.821\pm0.027\, (\Omega_m/0.25)^{0.4}$
\citep[see details in][]{lyapin18}.

Despite the fact that different samples of clusters of galaxies are
used in the works cited above, these constraints can not be considered
as completely independent ones, since very similar cluster mass scale
calibrations are used. Therefore, in our work below we use only the
constraints based on X-ray selected galaxy clusters sample from
\cite{av09a,av09b}, taking in account new cluster mass scale
calibrations \citep{lyapin18} --- these data are referred to as
\emph{V09\_H15} below.

\subsection{Weak gravitational lensing}

The measurements of weak gravitational lensing of distant galaxies in
deep optical surveys allow to measure inhomogeneity of matter
distribution that light passes, which in turn allow to obtain almost
direct measurement of matter density perturbation amplitude. The left
panel in the middle row of Fig.~\ref{fig:s8-om-all-sep} shows the
$\sigma_8$ constraints from the data of Canada-France Hawaii Telescope
Lensing Survey \citep[\emph{CFHTLenS},][]{heymans13}, and also the
constraints from the Dark Energy Survey first data release
\citep[\emph{DES},][]{des_gwl_y1_17}.

\subsection{Matter power spectrum and redshift space distortions}

The data from large spectroscopic sky surveys allow to obtain the
constraints on density perturbations amplitude from the matter power
spectrum recovered from the measurements of galaxy power spectrum, and
also studying the anisotropy of galaxies distribution in space with
redshifts as radial coordinate, which occur due to peculiar motion of
galaxies in disturbed gravitational field. Since these constraints are
based on the same data, in many cases they are combined with each
other and also with the constraints based on baryon acoustic
oscillations (BAO), which are also measured with these data and which
will be discussed below.

Right panel in middle row of Fig.~\ref{fig:s8-om-all-sep} shows the
constraints on density perturbations amplitude from the data on power
spectrum of galaxies in SDSS 4-th data release
\citep[\emph{MPK\_SDSS},][]{tegmark06}. In lower row of
Fig.~\ref{fig:s8-om-all-sep} the constraints from the measurements of
power spectrum of galaxies in \emph{WiggleZ} survey \citep[left panel,
\emph{MPK\_WiggleZ},][]{blake11,parkinson12}, and also redshift space
distortions measured in SDSS 12-th data release \citep[right panel,
\emph{RSD\_SDSS},][]{alam17} are shown. Note, that the constrains from
two last datasets contain also the results of baryon acoustic
oscillations measurements obtained from the same data.

\subsection{Combined constraints}

% In Fig.~\ref{fig:s8-om-all-sep} the constraints on density
% perturbations amplitude from \emph{Planck} CMB temperature anisotropy
% measurements are compared to the other constraints, discussed
% above. Note that these constraints are obtained from various
% cosmological datasets, using different methods, which can have
% absolutely different systematics. One can see that all these
% constraints give notably lower values of $\sigma_8$, as compared to
% \emph{Planck} TT constraints.

In the left panel of Fig.~\ref{fig:s8-om-all} all the constraints
discussed above are shown. All these data are in good agreement with
each other, with the exception of the data on Planck CMB temperature
anisotropy spectrum at high multipoles, $\ell>1000$, which give
significantly different $\sigma_8$ and $\Omega_m$ constraints.
% The left panel of Fig.~\ref{fig:s8-om-all} shows the constraints
% obtained as a combination of all the remaining data, excluding also
% the data on \emph{WiggleZ} survey and the data on the latest redshift
% space distortions measurements, because the data contain also the
% results of BAO measurements (see discussion below).
The combined constraints shown in the right panel of
Fig.~\ref{fig:s8-om-all} are obtained as a combination of the
following data:

\begin{itemize}
\item \emph{Planck} survey CMB temperature anisotropy spectrum at
  multipoles $\ell<1000$ (\emph{TT,$\ell<1000$});
  
\item \emph{Planck} LFI CMB polarization data at low multipoles,
  $\ell<30$ (\emph{lowTEB});
  
\item \emph{Planck} HFI CMB polarization data at low multipoles,
  $\ell<30$, used as a prior for reionization optical depth
  $\tau=0.055\pm0.009$ ($\tau0.055$);

\item \emph{Planck} CMB gravitational lensing potential measurements
  ($C^{\phi\phi}$);

\item South Pole Telescope CMB polarization measurements (\emph{SPTpol});
  
\item galaxy cluster mass function measurements from X-ray selected
  clusters (\emph{V09\_H15});
  
\item weak gravitational lensing of distant galaxies in Canada-France
  Hawaii Telescope Lensing Survey (\emph{CFHTLenS});
  
\item weak gravitational lensing of distant galaxies in Dark Energy
  Survey, first data release (\emph{DES});

\item luminous red galaxies power spectrum from Sloan Digital Sky
  Survey, data release 4 (\emph{MPK\_SDSS}).
\end{itemize}

The data on galaxies power spectrum in \emph{WiggleZ} survey and the
redshift space distortions data are not used here, because these data
contain also the results of BAO measurements, which will be discussed
below. In total, we use nine independent datasets, which are in good
agreement among each other. The common characteristic of these
datasets is that they produce direct constraints on matter density
perturbations amplitude, as it is discussed above.
%Along with that
Note, that almost all available data of that kind are used here,
\emph{excluding} the data on \emph{Planck} CMB temperature anisotropy
spectrum at high multipoles, $\ell>1000$. This combined dataset is
referred as \s8data\ below.

%{tab:s8data}

\begin{table}
  
  \caption{The constraints on cosmological parameters from the data on matter density perturbations measurements, $D(\sigma_8)$, in \emph{$\Lambda$CDM} model.} 
  \label{tab:s8data}
  \vskip 2mm
  \renewcommand{\arraystretch}{1.3}
  \renewcommand{\tabcolsep}{0.8cm}
  \centering
  \footnotesize
  \begin{tabular}{lll}
    \noalign{\doubleline}
    Parameter & value \\
    \noalign{\vskip 3pt\hrule\vskip 5pt}
    
    $ \Omega_b h^2            $ & $ 0.02271 \pm 0.00021  $  \\ %   1.45  0.63
    $ \Omega_c h^2            $ & $  0.1147 \pm  0.0012  $  \\ %   -2.52  0.82
    $ 100\theta_{MC}          $ & $ 1.04045 \pm 0.00060  $  \\ %   -0.78  0.50
    $ \tau                    $ & $  0.0592 \pm  0.0080  $  \\ %   -0.49  1.50
    $ {\rm{ln}}(10^{10} A_s)  $ & $   3.039 \pm   0.015  $  \\ %   -0.93  1.49
    $ n_s                     $ & $  0.9763 \pm  0.0057  $  \\ %   1.29  0.70
    $ H_0                     $ & $   69.37 \pm    0.61  $  \\ %   2.04  0.75
    $ \Omega_\Lambda          $ & $  0.7129 \pm  0.0071  $  \\ %   2.24  0.87
    $ \Omega_m                $ & $  0.2871 \pm  0.0071  $  \\ %   -2.24  0.87
    $ \sigma_8                $ & $  0.7921 \pm  0.0062  $  \\ %   -2.24  1.38
    $ z_{\rm re}              $ & $    7.99 \pm    0.77  $  \\ %   -0.60  1.51
    $ {\mbox{Age}}/{\mbox{Gyear}}   $ & $  13.754 \pm   0.034  $  \\ %   -1.00  0.60
    
    \noalign{\vskip 3pt\hrule\vskip 5pt}
  \end{tabular}
  
\end{table}

\begin{figure*}
  \centering
  \includegraphics[width=0.32\linewidth]{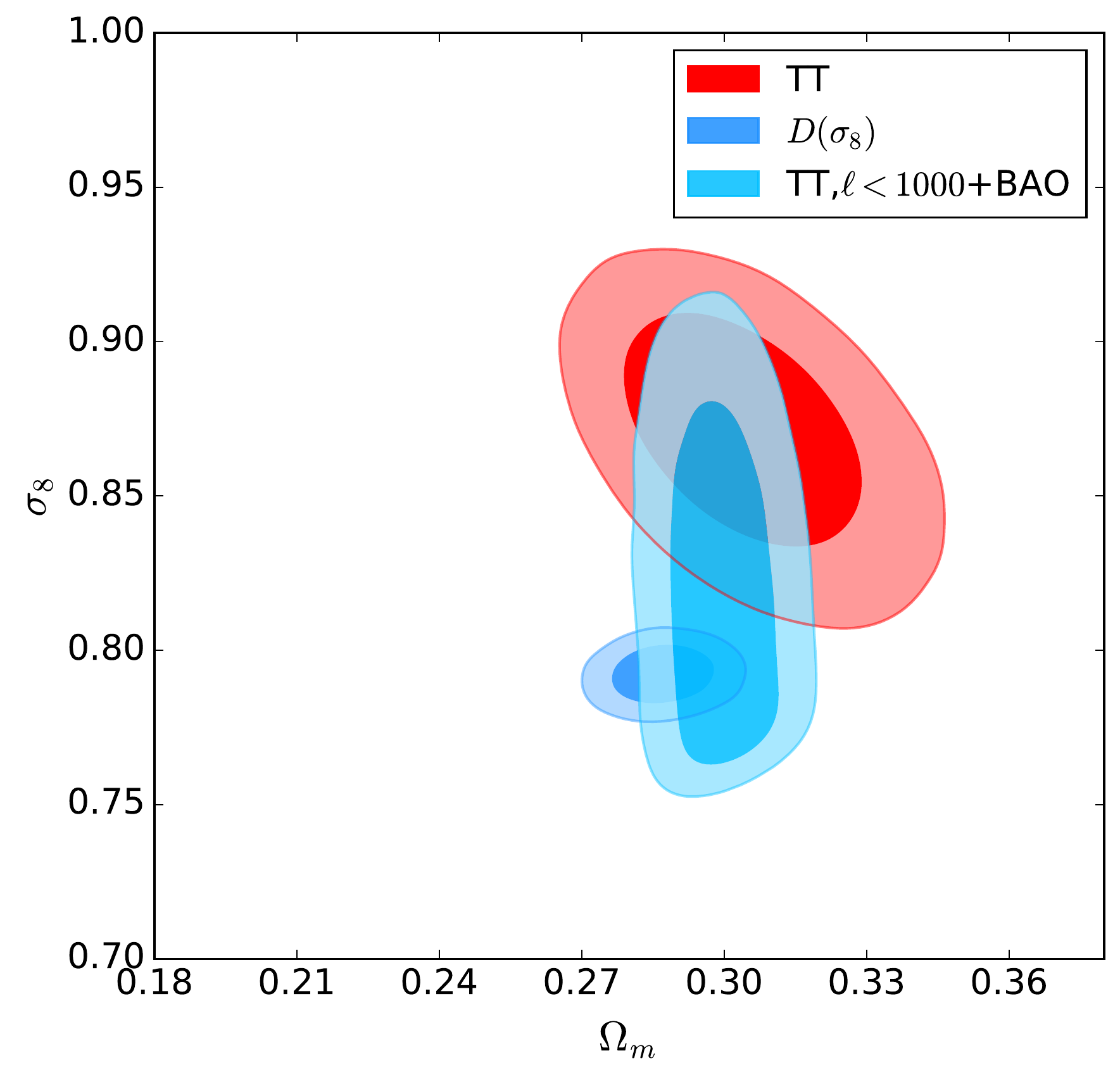} 
  \includegraphics[width=0.32\linewidth]{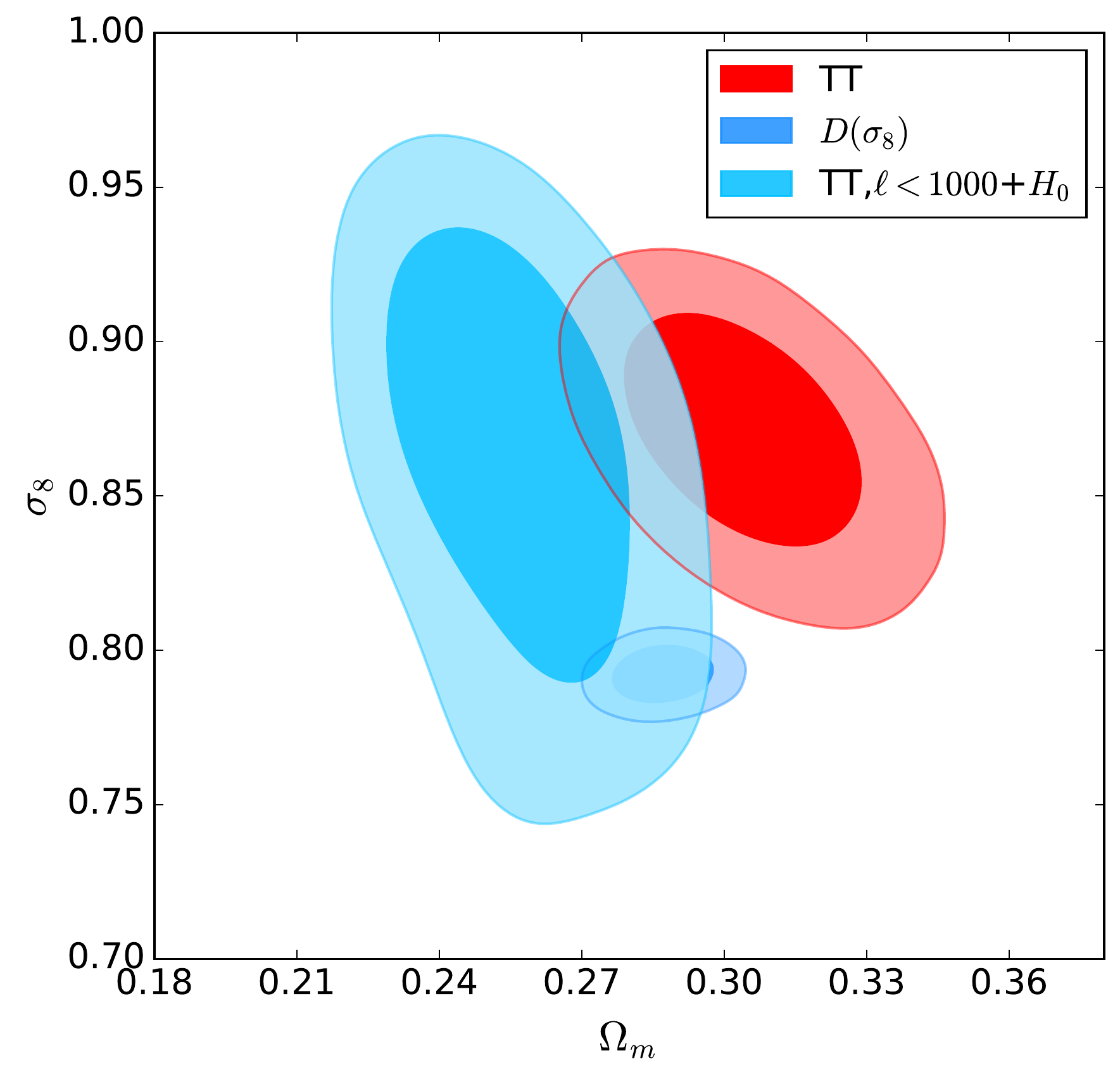}
  \includegraphics[width=0.32\linewidth]{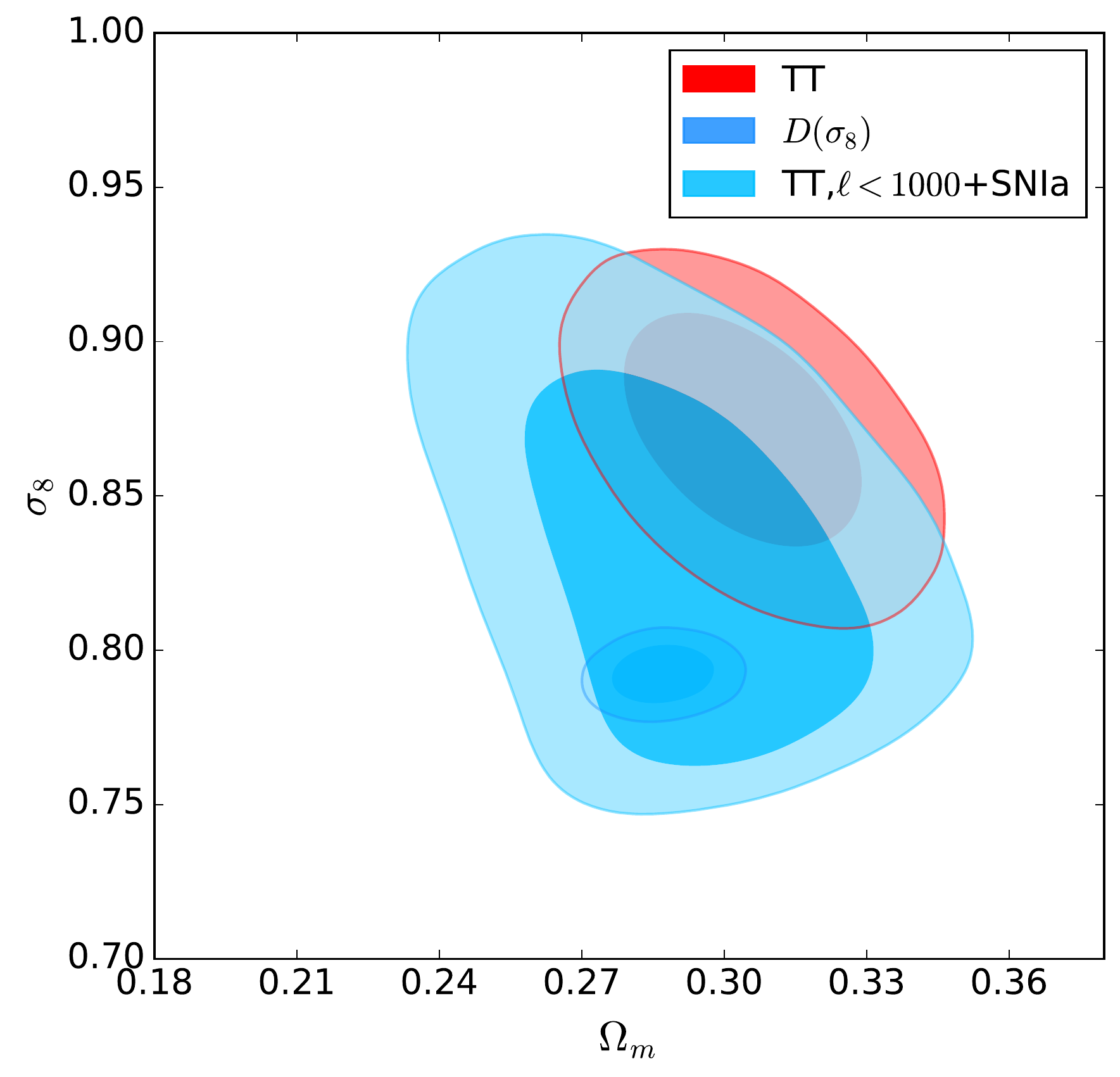}
  
  \caption{The constraints on mean matter density, $\Omega_m$, and
    linear density perturbations amplitude, $\sigma_8$, in
    \emph{$\Lambda$CDM} model from the \emph{Planck} CMB temperature
    anisotropy spectrum (\emph{TT}), from \s8data\ dataset, and also
    from the measurements of baryon acoustic oscillations (left
    panel), Hubble constant (middle panel) and type Ia supernovae at
    high redshifts (right panel), in combination with with
    \emph{Planck} CMB temperature anisotropy data at $\ell<1000$.}
  \label{fig:s8-om-geom} 
\end{figure*}

As it is shown in Fig.~\ref{fig:s8-om-all}, the $\sigma_8$ and
$\Omega_m$ constraints obtained with these data appears to be
significantly different, as compared to the constraints from
\emph{Planck} CMB temperature anisotropy spectrum at high multipoles,
$\ell>1000$, shown in Fig.~\ref{fig:s8-om-all} with red contours. The
\s8data\ dataset give the constraint
$\sigma_8\Omega_m^{0.25} = 0.580 \pm 0.006$, while \emph{Planck}
\emph{TT} data give $\sigma_8\Omega_m^{0.25} = 0.646\pm
0.017$. Therefore, the disagreement between these two datasets is
significant at about $3.7\sigma$ level.

The observed significance corresponds to only one deviation of such
amplitude in out of the order of 5000 realizations. In our case the
number of independent trials can be estimated as the number of
independent experiments, i.e. $\sim10$, and therefore this
disagreement should be considered as a significant one. Since in our
case one dataset (\emph{Planck} \emph{TT} spectrum at high multipoles)
contradicts to many other independent measurements, most probably, it
is these data which contains some unknown systematics. However, even
if this disagreement were explained by statistical reasons only, one
should discard significantly deviating data, since we are interested
in statistically best possible estimates of the parameters. The real
reasons of this disagreement should be studied further. However,
already now it is clear that the combination of these data
\emph{should not} be used currently to constrain cosmological
parameters.

The \s8data\ combined dataset consist of many different measurements
of density perturbation amplitude, which are found to be in good
agreement between each other. Therefore, the cosmological parameters
constraints based on this dataset should be reliable and they are
studied in detail below. The constraints on various cosmological
parameters from \s8data\ combined dataset are presented in
Table.~\ref{tab:s8data}. Note, that the accuracy of cosmological
parameters determination appears to be comparable to that obtained
from \emph{Planck} 2015 data release in combination with external data
\citep{PSZ2cosm}. However, best estimates of some parameters are
shifted for about $2\sigma$. For example, the best estimates of
$\sigma_8$ and $\Omega_m$ parameters are shifted to lower values,
while the best estimate of Hubble constant, $H_0$, appears to be
shifted high, as compared to \emph{Planck} 2015 constraints. This is
explained mostly by the fact, that the data on \emph{Planck} \emph{TT}
spectrum at high multipoles, $\ell>1000$, are not used in \s8data\
dataset.

Note, that as compared to \emph{Planck} constraints published earlier,
the constraints from \s8data\ dataset are in much better agreement
with the pre-\emph{Planck} constraints, which were obtained from the
combination of various cosmological data \citep{bv12,br13}. The
remaining $\approx2\sigma$ differences between $\Omega_m$ and $H_0$
measurements are explained by different galaxy cluster mass scale
calibration and by using the additional $H_0$ constraint from direct
Hubble constant measurements \citep{riess11} in the earlier works
cited above.

\section{The measurements of distances and expansion rate of Universe}

The data on baryon acoustic oscillations, type Ia supernovae and
direct measurements of Hubble constant can not be used to obtain
direct constraints on density perturbation amplitude. The constraints
on mean matter density, $\Omega_m$, and linear density perturbations
amplitude, $\sigma_8$, in \emph{$\Lambda$CDM} model from these data
are shown in Fig.~\ref{fig:s8-om-geom}. The BAO data are taken from
the measurements of luminous red galaxies power spectrum in SDSS 12-th
data release \citep{alam17}, power spectrum of less distant galaxies
measured with the data of SDSS 7-th data release \citep{ross15}, and
also from the power spectrum of nearby galaxies measured in \emph{6dF}
survey \citep{botler11}. The direct Hubble constant measurement,
$H_0 = 73.48\pm 1.66$, was taken from \cite{riess18}, the supernova
type Ia constraints are based on joint lightcurve analysis of
supernovae from SDSS and SNLS samples
\citep[\emph{JLA},][]{betoule14}.

One can see that these data are in good agreement with both the data
on \emph{Planck} \emph{TT} spectrum at high multipoles and with
\s8data\ dataset. There is only a weak, approximately $2\sigma$,
tension with the direct Hubble constant measurement. The constraints
from the \s8data\ dataset are located approximately in between the
constraints from BAO and direct measurement of Hubble constant. The
combination of these data gives different constraints on sum of
neutrino mass and number of relativistic species, which are discussed
in detail below.

\begin{figure*}
  \centering
  \includegraphics[width=0.32\linewidth]{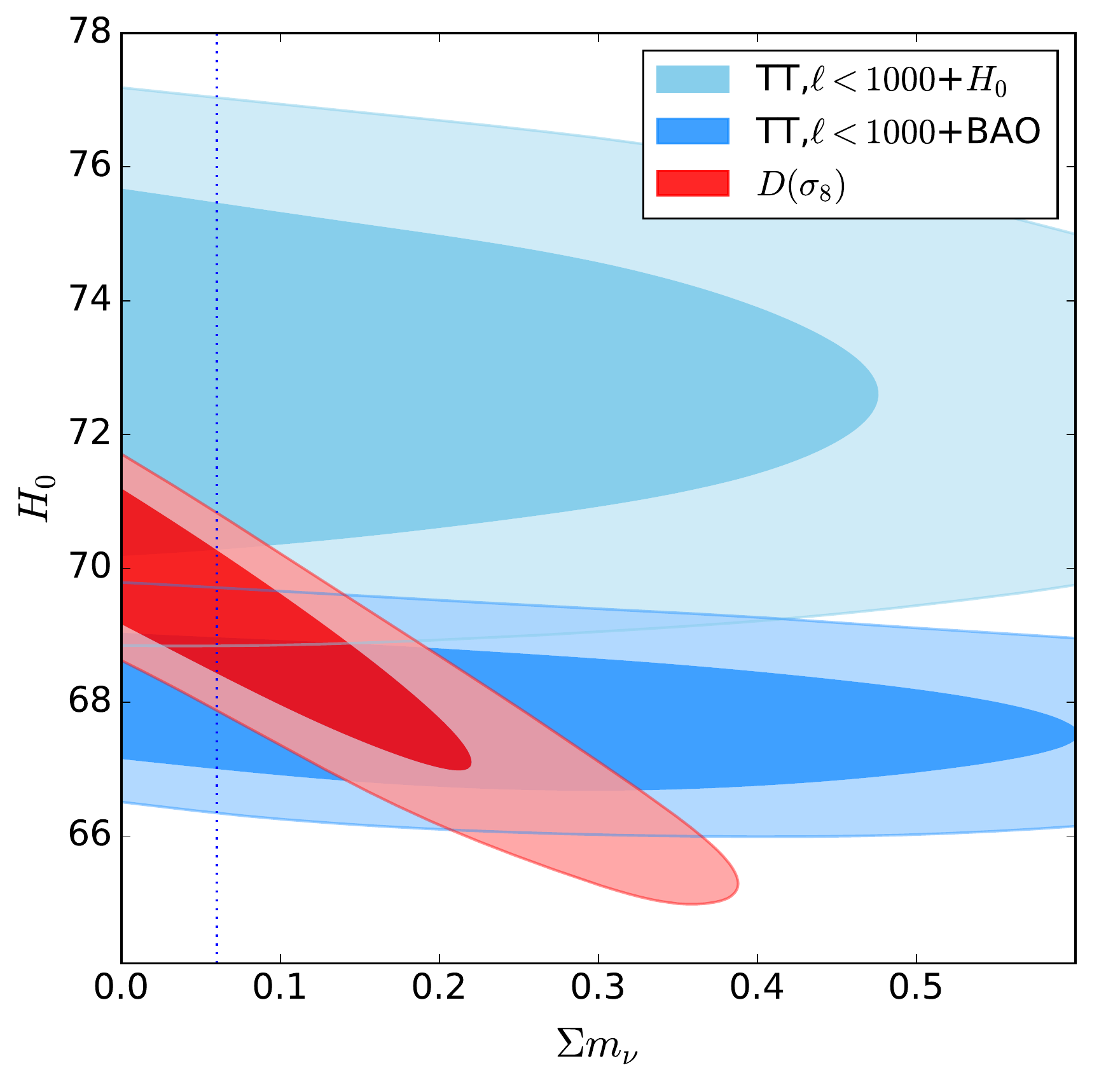} 
  \includegraphics[width=0.32\linewidth]{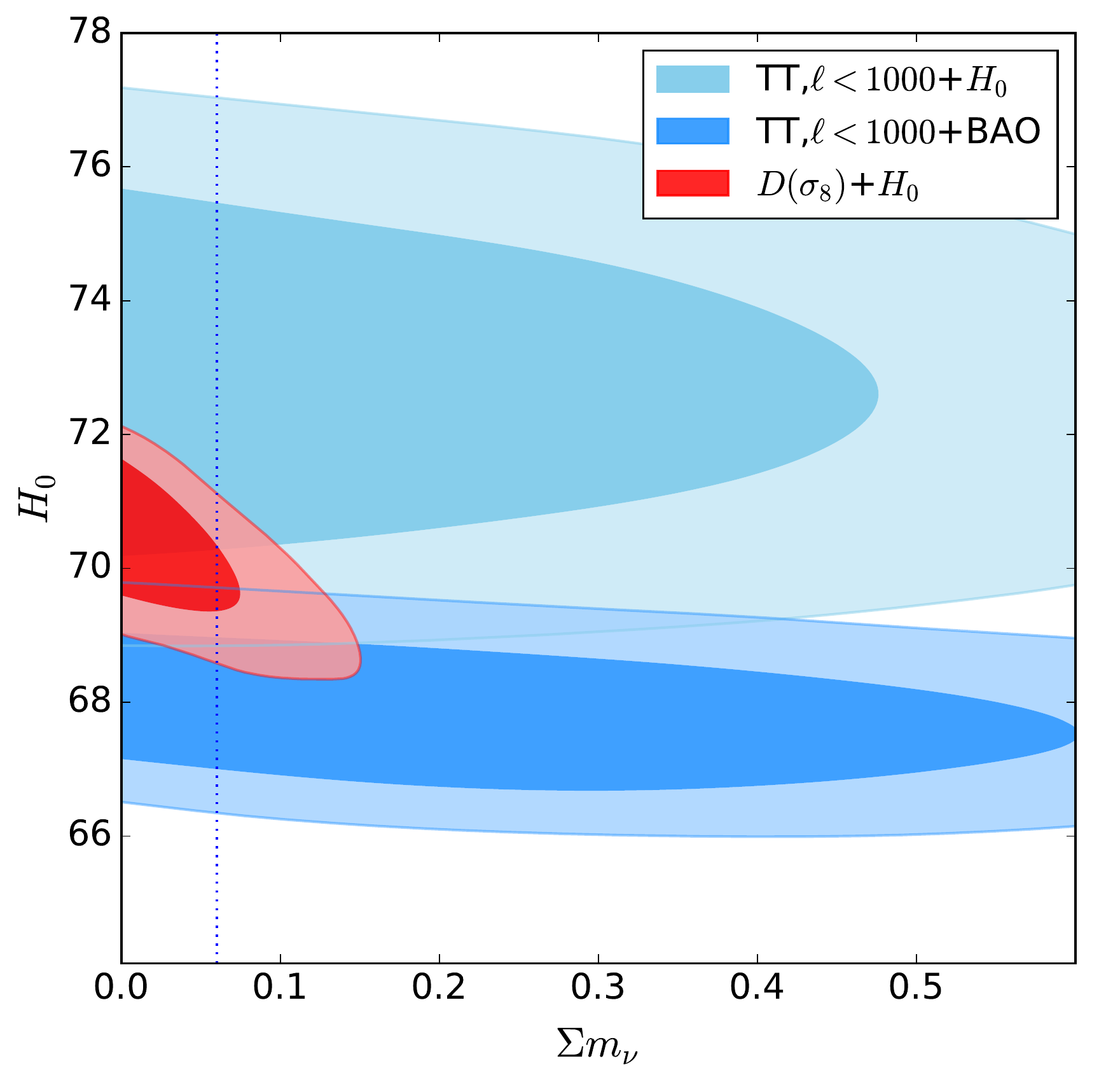}
  \includegraphics[width=0.32\linewidth]{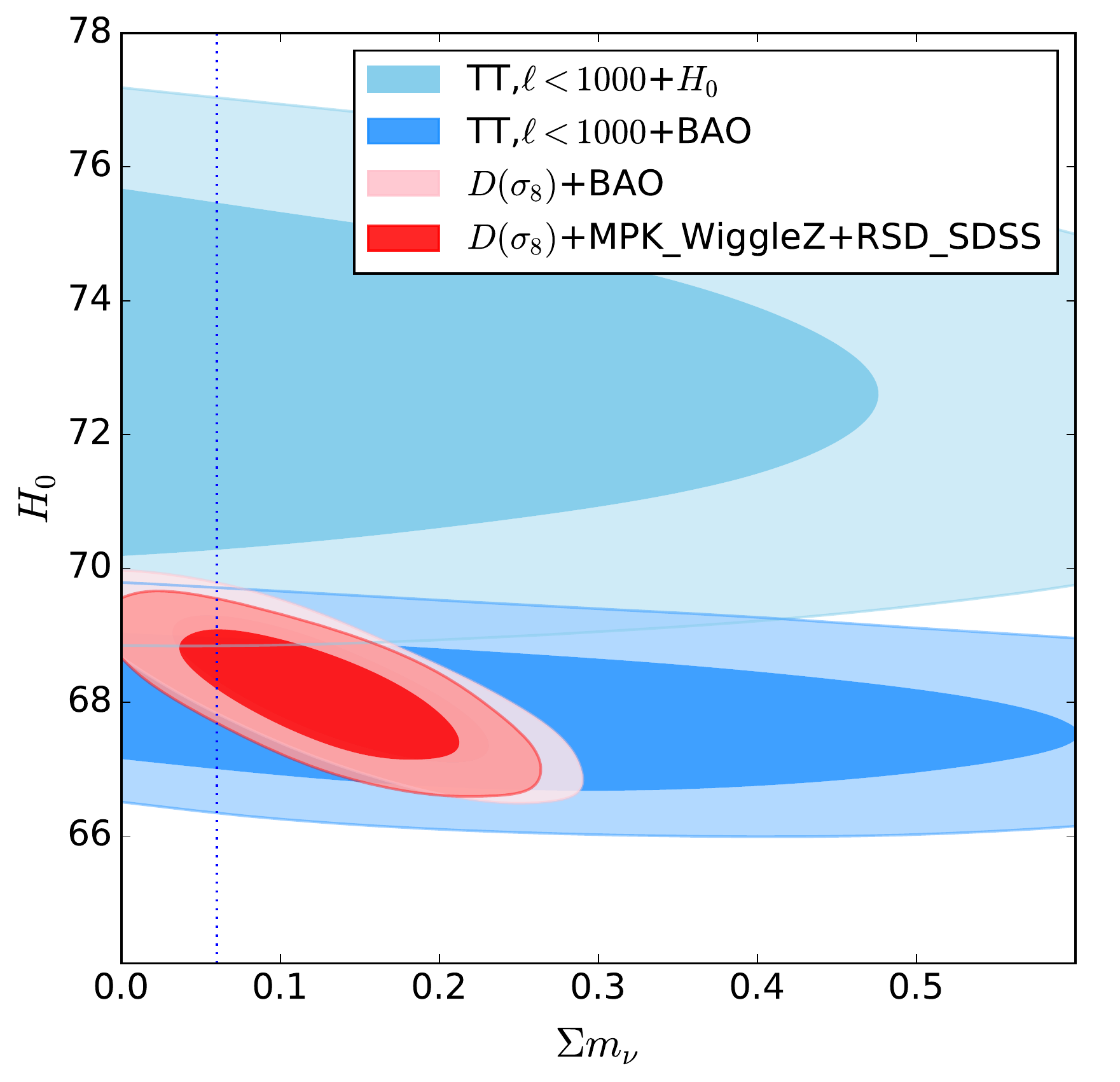}

  \caption{The constraints on sum of neutrino mass and Hubble constant
    from the BAO and direct $H_0$ data, and also from \s8data\ dataset
    in \emph{$\Lambda$CDM}+$m_\nu$ model. Vertical dotted line show
    the value $\Sigma m_\nu=0.06$~eV, which approximately corresponds
    to the minimum sum of neutrino mass from the observations of
    neutrino oscillations.}
  \label{fig:mnu-H0} 
\end{figure*}

\section{Constraints on sum of neutrino mass and number of
  relativistic species}

The constraints on sum of neutrino mass and Hubble constant from the
BAO and direct $H_0$ data, and also from \s8data\ dataset in
\emph{$\Lambda$CDM}+$m_\nu$ model are shown in Fig.~\ref{fig:mnu-H0}.
Vertical dotted line show the value $\Sigma m_\nu=0.06$~eV, which
approximately corresponds to the minimum sum of neutrino mass from the
observations of neutrino oscillations
\citep[e.g.,][]{LesgourguesPastor06}.  One can see, that $H_0$
constraints does not depend on sum of neutrino mass, as long as BAO
and $H_0$ data are considered, as expected. The \s8data\ dataset also
give rather good constraint on $H_0$, but this constraint depends
strongly on sum of neutrino mass. It can be shown that degeneracy is
caused, eventually, by the suppression of density perturbation growth
by massive neutrinos.

From Fig.~\ref{fig:mnu-H0} one can see that different combinations of
datasets can give different $\Sigma m_\nu$ constraints. So, for
example, \s8data\ taken separately produce the constraint:
$\Sigma m_\nu<0.308$~eV, while in combination with direct $H_0$
measurements this same dataset gives: $\Sigma m_\nu<0.117$~eV. Note,
that in the last case the upper limit (at 95\% confidence level) turns
out to be close to the lower limit for sum of neutrino mass in the
case of inverse mass hierarchy, $\Sigma m_\nu\approx 0.1$~eV
\citep[e.g.,][]{LesgourguesPastor06}.

\begin{figure}
  \centering
  \includegraphics[width=0.64\linewidth]{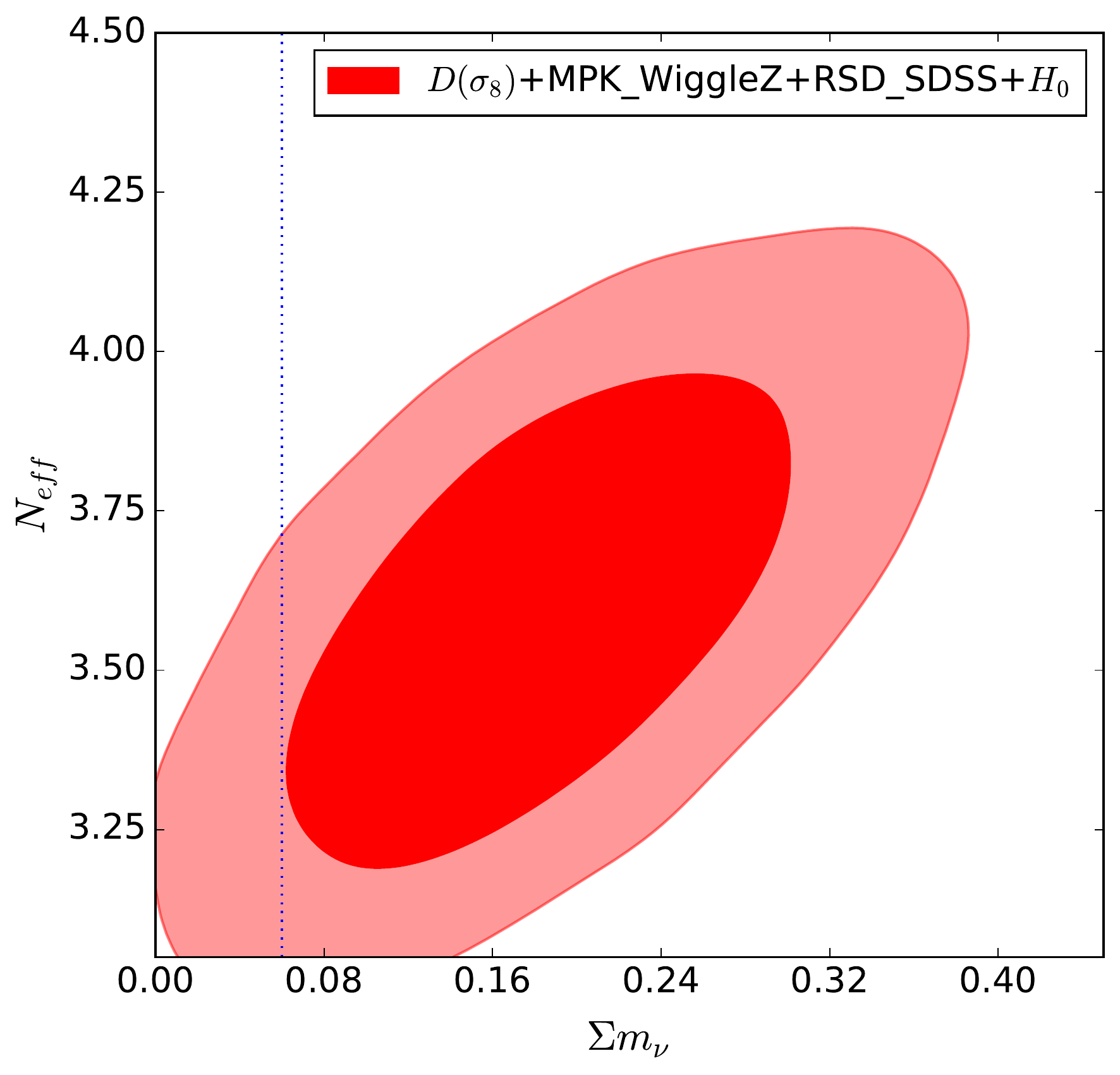}

  \caption{The constraints on sum of neutrino mass, $\Sigma m_\nu$,
    and number of relativistic species, \neff, in
    \emph{$\Lambda$CDM}+$m_\nu$+$\neff$ model, from the
    \s8data+\emph{MPK\_WiggleZ}+\emph{RSD\_SDSS}+$H_0$ combined
    dataset.}
  \label{fig:mnu-nnu} 
\end{figure}

In combination with the BAO data the \s8data\ dataset give the
measurement of non-zero sum of neutrino mass, at approximately
$2\sigma$ significance: $\Sigma m_\nu=0.128\pm0.056$~eV, the 95\%
upper limit in this case is $\Sigma m_\nu<0.233$~eV. However, the
measurement of Hubble constant based on BAO observations is known to
be model dependent. This measurement is obtained using the sound
horizon size as a standard ruler and therefore depends on the
relativistic energy density in early Universe, which is usually
parameterized through the effective number of relativistic species,
\neff. Therefore, if one add \neff\ as a free parameter to
cosmological model, one could reconcile all the data discussed above.
The $\Sigma m_\nu$ and \neff\ constraints in that model, using all the
cosmological data discussed above
(\s8data+\emph{MPK\_WiggleZ}+\emph{RSD\_SDSS}+$H_0$), are:
$\Sigma m_\nu=0.189\pm0.080$~eV and $\neff = 3.61\pm0.25$. These
constraints are shown in Fig.~\ref{fig:mnu-nnu}.

Similar measurements for non-zero sum of neutrino mass and larger than
standard number of relativistic species were obtained earlier when the
data on galaxy cluster mass function were combined with other
cosmological data, before the publication of the first \emph{Planck}
survey results \citep{bv12,br13,hou14}. These constraints, formally do
not contradict to the constraints presented in Fig.~\ref{fig:mnu-nnu},
however, in the last case allowed region of parameter space turns out
to be much smaller in its size and is shifted closer to the standard
values of \emph{$\Lambda$CDM} model.

The remaining deviations of $\Sigma m_\nu$ and \neff\ from
$\Lambda$CDM standard values are driven by the remaining tensions
between the direct $H_0$ measurements and BAO observations. If in
future BAO would measure higher values of $H_0$, consistent with
current direct $H_0$ measurements, both $\Sigma m_\nu$ and \neff\
combined constraints will be consistent with $\Lambda$CDM standard
values. If instead direct $H_0$ measurements would produce lower
values of $H_0$, consistent with current BAO measurements, standard
\neff\ and non-zero $\Sigma m_\nu$ will be measured with the combined
data.

\section{Discussion and conclusions}

In this work various methods of density perturbations amplitude
measurements, which use both Planck CMB anisotropy data and other
cosmological data, are compared. We consider the measurements based on
the CMB temperature and polarization anisotropy spectra, CMB lensing
potential, galaxy cluster mass function, weak gravitational lensing of
distant galaxies, matter power spectrum and redshift space distortions
in spectroscopic surveys of galaxies. All these measurements were
carried out in different experiments, are completely independent and
have different systematic uncertainties.

It turns out that all these measurements are in good agreement with
each other, \emph{with the exception} of the data on Planck CMB
temperature anisotropy spectrum at high multipoles, $\ell>1000$.  This
measurement contradict to all other constraints obtained both from
remaining Planck CMB anisotropy data and from other cosmological data,
at about $3.7\sigma$ significance level. The true explanation of this
disagreement should be identified in a separate study. However,
already now it is clear that the combination of these data
\emph{should not} be used currently to constrain cosmological
parameters.

Since in our case one dataset (\emph{Planck} TT spectrum at high
multipoles, $\ell>1000$) contradicts to large number of other
independent measurements, it is in these data some unknown systematics
may be contained. These data can not be compared to \emph{WMAP} all
sky survey data due to lower angular resolution of \emph{WMAP}
\citep{wmap9yr}. However, it can be shown that at larger angular
scales, i.e.\ at multipoles about $\ell<1000$, CMB temperature
anisotropy measurements made by \emph{Planck} and \emph{WMAP} are in
good agreement \citep{huang18}.

The possibility of the presence of unknown systematical uncertainties
in \emph{Planck} survey temperature anisotropy data was studied
through the comparison of these data with the temperature anisotropy
data from South Pole Telescope \citep[SPT,][]{aylor17,hou18} and
Atacama Cosmology Telescope \citep[ACT,][]{louis14,louis17}. It was
shown that there is a good agreement between these data inside the SPT
and ACT footprints. SPT cosmological parameters constraints were
compared to those from the combined \emph{Planck} temperature and
low-$\ell$ polarization all sky data and weak $2\sigma$ tensions
between these datasets were found --- somewhat higher value of $H_0$
and lower value of $\Omega_m$ were measured by SPT \citep{aylor17}, in
agreement with the results discussed in our work.

Various measurements of matter density perturbation amplitude,
excluding \emph{Planck} CMB temperature anisotropy data at high
multipoles, $\ell>1000$, combined together, are denoted as \s8data\
dataset above.  Since this dataset consist of many different
measurements of density perturbation amplitude, which are found to be
in good agreement between each other, the cosmological parameters
constraints based on this dataset should be reliable and they are
studied in detail.

The \s8data\ dataset produce the constraints with best estimates of
some parameters shifted by about $2\sigma$ as compared to
\emph{Planck} 2015 constraints \citep{planck15_cosm}. For example, the
constraints on $\sigma_8$ and $\Omega_m$ are shifted to lower values,
but the Hubble constant constraint appears to be shifted
high. Therefore, if \emph{Planck} CMB temperature anisotropy spectrum
data at high multipoles, $\ell>1000$, are excluded, the tension in
$H_0$ constraints between \emph{Planck}+BAO data and direct $H_0$
measurements \citep[e.g.,][]{PSZ2cosm} is weaken, and the tension in
$\sigma_8$ measurements between \emph{Planck} and large scale
structure data
\citep[e.g.,][]{PSZ2cosm,planck15_cosm,planck15_par_shifts} disappear
completely. In addition, as it is shown above, the gravitational
lensing amplitude anomaly observed in \emph{Planck} CMB data also
disappear.

% The combined dataset on the density perturbations amplitude
% measurements, \emph{excluding} \emph{Planck} CMB temperature
% anisotropy spectrum data at high multipoles, $\ell>1000$, which
% reffered as \s8data\ above, consists of a large set of heterogeneous
% data which are in good agreement between each other. 

The combined \s8data\ dataset give the Hubble constant measurement,
which is located approximately in between the constraints from the BAO
and direct $H_0$ measurements. From the different combination of these
data one can obtain different constraints on sum of neutrino mass. All
these data can be reconciled completely, if one assume non-zero sum of
neutrino mass and larger than standard number of relativistic species.

Qualitatively, similar constraints were obtained earlier when the data
on galaxy cluster mass function were combined with other cosmological
data before the release of first \emph{Planck} survey results
\citep{bv12,br13,hou14}. However, according to new constraints,
allowed region of parameter space appears to be much smaller in size
and shifted closer to the standard values of \emph{$\Lambda$CDM}
model. The remaining deviations of $\Sigma m_\nu$ and \neff\ from
$\Lambda$CDM standard values are driven by the remaining tensions
between the direct $H_0$ measurements and BAO observations.

\acknowledgements

The work is supported by Russian Science Foundation grant 18-12-00520.
This project was carried out using the computational resources of
Spectrum-Roentgen-Gamma project data archive at IKI and also the
resources of MVS-10P computer cluster at Joint Supercomputer Center of
the Russian Academy of Sciences (JSCC RAS).

\end{document}